\documentclass[11pt]{article}
\usepackage[margin=1.1in]{geometry}

\usepackage{amssymb}
\usepackage{amsmath}
\usepackage{tikz}
\usetikzlibrary{arrows,calc,shapes,decorations.pathreplacing}
\usepackage[driverfallback=dvipdfm]{hyperref}
\hypersetup{pdftex,colorlinks=true,allcolors=blue}
\usepackage{caption}
\usepackage{subcaption}

\usepackage{algorithm}
\usepackage{algorithmicx}
\usepackage{algpseudocode}

\usepackage{color}

\newtheorem{theorem}{Theorem}

\newtheorem{proposition}[theorem]{Proposition}

\newtheorem{definition}{Definition}

\newcommand*\diff{\mathop{}\!\mathrm{d}}
\def\P{\mathrm{P}}
\def\E{\mathrm{E}}

\def\l{\lambda}

\def\m{\mu}

\def\D{\Delta}

\def \x{\xi}

\def \a{\alpha}

\def \b{\beta}

\DeclareMathOperator*{\argmin}{arg\,min}

\begin{document}

\title{A survey of queueing systems with strategic timing of arrivals\footnote{To appear in Queueing systems: Theory and Applications.}}

\author{Moshe Haviv\footnote{Deaprtment of Statistics and Data Science and the Center for Rationality, The Hebrew University of Jerusalem and School of Data Science, The Chinese University of Hong Kong, Shenzhen Campus.} and Liron Ravner\footnote{Deaprtment of Statistics, University of Haifa, and Korteweg-de Vries Institute for Mathematics, University of Amsterdam.}}

\date{\today}
\maketitle

\begin{abstract}
Consider a population of customers each of which needs to decide independently when to arrive to a facility that provides a service during a fixed period of time, say a day. This is a common scenario in many service systems such as a bank, lunch at a cafeteria, music concert, flight check-in and many others. High demand for service at a specific time leads to congestion that comes at a cost, e.g., for waiting, earliness or tardiness. Queueing Theory provides tools for the analysis of the waiting times and associated costs.  If customers have the option of deciding when to join the queue, they will face a decision dilemma of when to arrive. The level of congestion one suffers from  depends on others behavior and not only that of the individual under consideration. This fact leads customers to make strategic decisions regarding their time of arrival. In addition, multiple decision makers that affect each other's expected congestion,  call for non-cooperative game theoretic analysis of this strategic interaction. This common daily scenario has prompted a research stream pioneered by the ?/M/1 model of Glazer and Hassin \cite{GH1983} that first characterized an arrival process to a queue as a Nash equilibrium solution of a game. This survey provides an overview of the main results and developments in the literature on queueing systems with strategic timing of arrivals. Another issue is that of social optimality, namely the strategy profile used by customers that optimizes their aggregate utility. In particular, we review results concerning the price of anarchy (PoA), which is the ratio between the socially optimal and the equilibrium utilities. 
\end{abstract}

\section{Introduction}\label{sec:intro}

The choice of when to join a congested service system is a common daily scenario for most people. This is the case when a number of customers need to decide when to arrive to a bank on a busy Monday. One can go  early to `avoid the crowd', perhaps even queueing before the opening, but this entails waking up early. Another option is to arrive later in the day but then the queue may be long and potentially being late for work.  Another example is when passengers who are booked for an 11pm flight need to decide when to arrive to the check-in counter, that opens at 7pm. Imagine what will happen if all  will follow the airline advice to arrive at least three hours prior to boarding. Ideally, travelers want to arrive as late as possible, say at 10pm, but delaying the arrival for too long may lead to waiting a long time for the other travelers that are already in the check-in queue. Also in this case there is a very big cost associated with being too late, namely missing the flight. This tradeoff calls for strategic decision making by all arriving customers. The game-theoretic concept of a Nash equilibrium provides a  natural model for this situation because it prescribes an arrival process such that the cost for all arriving customers is constant over the relevant period. This survey provides an overview of the research focusing on the modeling and analysis of queueing systems with the above strategic interaction.

Many queueing models assume an ongoing stream of joining customers whose number is potentially infinite as time passes and therefore resulting in an ongoing queueing process with some steady-state distribution. However, temporal decision making requires a queueing model with a finite number of customers and a finite time duration; we therefore consider a finite population of customers that require service from a server that operates during a specified time period. The customers individually decide when to arrive at the system, and if it is busy they join a queue and wait for their turn to be served.  As discussed above, the scenario of a finite population of customers arriving and being served during a finite period is common in queueing applications, however, it is less common in the theoretical analysis of queueing systems. One key difference from standard queueing models is that the process is inherently transient and the traditional steady-state analysis is not applicable. Another is that the arrival process is endogenous in the sense that it is a function of the economic parameters associated with the customers and, more crucially, of the strategic interaction which typically yields a non-homogeneous in time arrival process with dependent arrival epochs, i.e., not a renewal process. This process stems from the order statistics of the individual arrival-time decisions. 

We first provide a general model description that will enable using, as much as possible, the same consistent notation throughout the paper, rather than the specific notations used in the discussed papers. Suppose a population of $N<\infty$ customers require service, where $N$ could be a constant, random variable or even a fluid volume. Each customer $i$ has a service requirement $X_i$ and for now we assume that customers are statistically homogeneous; the service requirements are independent and identically distributed random variables. Customers can arrive during  an admission period $\mathcal{T}$ that can be a continuous interval, e.g., $[0,T]$ or $(-\infty,\infty)$, or a discrete grid $\{t_0,t_1,\ldots\}$. Services are provided by a single server or a number of servers that operate during some time interval $\mathcal{S}$. For example $\mathcal{S}=(-\infty,\infty)$ is a system that is always ``on" like a road and $\mathcal{S}=[0,T]$ is a system with opening and closing times such as a bank. In some cases queueing may also be allowed before service period initiation and/or service may continue after the admission period ends. Customer~$i$ can choose his/her arrival time $\tau_i$ and so a mixed strategy defines a probability distribution on $\mathcal{T}$ with cdf $F_i$. Customers wish to minimize the expectation of a cost function $C_i(\tau_1,\ldots,\tau_N;\Theta)$ that depends on the random arrival times of all other customers and a set of parameters $\Theta$ such as service rate and delay penalty. Note that $N$ can be a random variable and therefore $C$ is a function from the set of arbitrary sized sequences to the real line. Naturally, customer~$i$ has discretion only over $\tau_i$ and does not know the realization of $\tau_{-i}$ (the set of arrival times of customers excluding $i$), let alone have control on it. The specific form of the function is determined by the specific queueing system parameters, e.g., number of servers or service discipline, together with the distribution of the arrival process which is determined by the arrival strategies applied by all customers. 

A Nash equilibrium is a profile of strategies $\tau_i\sim F_i$, $1 \leq i \leq N$, such that any customer $i$ obtains a minimal expected cost $F_i \in \arg\min_{F}\E [C_i(\tau_1,\ldots \tau,\ldots,\tau_N)]$, where $\tau$ induces distribution~$F$. The majority of the research to date is focused on homogeneous customers and symmetric equilibrium solutions such that all customers use the same arrival strategy $F$. In many cases a symmetric equilibrium is the only possible outcome and in a sense it is a natural solution concept to adopt when the population is large and anonymous as it does not require identification of individual customers. However, we will mention some cases where asymmetric equilibria exist, even when customers are homogeneous. 

Any strategy profile used by customers results in some social cost which is defined as the sum of the costs incurred by the individual participants. It is natural to look for a profile which minimizes this cost. This criterion is different than the Nash equilibrium criterion, which usually leads to higher costs. The ratio between the corresponding costs, or the worst case when the Nash equilibrium is not unique, is called the price of anarchy (PoA).

We will first present two simple examples that illustrate some of the basic properties of the equilibrium and the socially optimal solutions.
	
\noindent{\bf Example 1.} Our first example and the pioneering model of this field is the ?/M/1 model introduced by Glazer and Hassin \cite{GH1983}. The population size $N$ is assumed to follow a Poisson distribution with mean $\lambda$. The service times are independent and exponentially distributed with rate $\mu$, and service is obtained from a single server that admits customers during a specified period of time $[0,T]$. Customers are served on a first come first served basis and can queue up before the opening time, i.e., $\mathcal{T}=(-\infty,T]$. All customers that arrive up to, and including, time $T$ are served ($\mathcal{S}=[0,\infty)$), in other words the server will continue working after $T$, if needed, until the system is empty. The objective of the customers is to minimize their sojourn time in the system. If all customers arrive according to a symmetric continuous arrival strategy $F$ with density $f$ on an interval $[t_a,T]$, then the arrival process to the queue $A(t)$ is a non-homogeneous in time Poisson process with rate $\lambda(t)=\lambda f(t)$. Therefore, in this case the queueing process is an $\mathrm{M}_t$/M/1. Interestingly, under reasonable axiomatic assumptions, a singled out arriving customer believes that the number of other customers arriving follows the same distribution: Poisson with mean $\lambda$. This is a special property of the Poisson distribution, but for general distributions the conditional distribution of the number of other arriving customers can be derived in a straightforward manner.  We elaborate in Section~\ref{sec:prior} on how the (prior) distribution of the number of customers changes conditional on the event that a tagged customer is actually a member of the randomly sized population.

In~\cite{GH1983} it is shown that the Nash equilibrium arrival strategy is given by a uniform distribution $f(t)=c$, where $c>0$ is a constant, on $t\in[t_a,0]$ where $t_a<0$, and a density $f(t)$ that solves a system of functional differential equations for $t\in[0,T]$. We first provide here an outline of the method for obtaining this solution, and later in Section \ref{sec:equilibrium} go into more detail for a generalization of this model. The equilibrium condition dictates that the waiting cost equals some constant $w$ on $[t_a,T]$, before the opening instant this condition translates to
\[
\E\left[\frac{A(t)+1}{\mu}\right]- t=w, \ t\in[t_a,0),
\]
and after the opening to 
\[
\E\left[\frac{Q(t)+1}{\mu}\right]=w, \ t\in[0,T],
\]
where $A(t)$ is the number of arrivals until time $t$ and where $Q(t)$ denotes the number of customers in the system at time $t$. The first condition is easily solvable because $\E[A(t)]=\lambda F(t)$, while the second already entails considering the dynamics of the queueing process $Q(t)$. This can be done by considering the transient state probabilities given by the Kolmogorov backward equations. In principle, the above can be solved for any $t_a$, where   $t_a$ can be found by a search procedure with termination condition $F(T)=1$. 

From a social welfare point of view, the equilibrium outcome is clearly not optimal. This is because a positive mass of customers arrive before the opening and wait while the server is not doing any work. Any such arrival profile can be improved by moving this mass to time zero, i.e., setting $\tilde{F}(t)=0$ for any $t<0$ and a point mass at $t=0$: $\tilde{F}(0)=F(0)$. Indeed, Hassin and Kleiner \cite{HK2011} show that the socially optimal strategy has a positive mass at zero, followed by a continuous distribution on $(0,T)$ and a positive mass at $T$. We will discuss the details of social optimization in Section \ref{sec:optim}. \hfill $\Diamond$

Finding the symmetric equilibrium arrival distribution $F$ for different settings is the goal of most of the literature discussed in this survey. As is already evident in the ?/M/1 example, this, at least numerically, is typically not a trivial task. The issues of existence and uniqueness of Nash equilibria need to be dealt with carefully and also devising reasonable methods for numerically approximating the equilibrium is often challenging. The combination of transient analysis together with elaborate time and service-regime dependence, makes each model challenging in a unique way and calls for specialized solutions, and there is still much work to be done for better understanding of these models. The following example presents a special case where $F$ can be explicitly derived for a game with just two customers. This example further illustrates that asymmetric equilibria may exist as well.

\vspace{0.2cm}
\noindent{\bf Example 2.} Two customers seek independent and exponentially distributed service with parameter $\mu$ from a server who opens his service at time $0$. Customers have to select an arrival time which needs to be in the interval $\mathcal{T}=[0,T]$. The customers complete their service even if it goes beyond time $T$, i.e., $\mathcal{S}=[0,\infty)$. Both wish to minimize their own waiting time, assuming the service regime of first-come first-served and ties are broken randomly. Suppose customer $1$ arrives at time $0$, then it is easy to verify that arriving at $T$ is optimal for customer $2$ if $T\geq \log2/\mu$. Therefore, in this case an asymmetric equilibrium comes with one of them arriving at time $0$ and the other at time $T$, and hence two pure equilibria exist. For $T> \log2/\mu$ an additional symmetric equilibrium exists such that both customers use the following mixed strategy: arrive at time~$0$ with probability $2/(2+\mu T-\log 2)$, no arrivals during the time interval $(0,\log2/\mu)$, and a uniform density along the time interval $[\log 2/\mu,T]$. If $T< \log2/\mu$ then both arriving at time $0$ is the unique symmetric equilibrium. 

The socially optimal solution is trivially obtained by one customer arriving at $0$ and the other at $T$. The social cost in this case is $e^{-\mu T}/\mu$. Thererfore, if $T\geq \log2/\mu$ the pure strategy equilibria described above are also socially optimal. The social cost for the symmetric equilibrium for both customers arriving at $0$ is $1/\mu$ and for the symmetric mixed equilibrium detailed above it is $(\mu(2+\mu T-\log 2))^{-1}$. In both of the latter cases the social cost in equilibrium is strictly  higher than the optimal cost. More details on this can be found in~\cite{HR2015}. $\Diamond$

\subsection{Paper organization}\label{sec:organization}

This survey is organized as follows. Firstly, Section~\ref{sec:model} provides an in depth discussion of the model. Specifically, Section~\ref{sec:queue} introduces the queueing model associated with a finite population of customers randomizing their arrival times independently. Section~\ref{sec:prior} provides the assumptions and definitions required for analysing a game with a random population of customers, and in particular the significance of assuming that the size of the population is a Poisson random variable. Section~\ref{sec:game} defines the game of timing arrivals and the concept of a Nash equilibrium arrival profile and how it relates to the queue dynamics. Section~\ref{sec:equilibrium} reviews the main results regarding the Nash equilibrium solutions for various models. Moreover, the methods for computation of the equilibrium are reviewed along with the difficulties that arise when moving away from the FCFS single server model. Section~\ref{sec:optim} surveys the issue of system optimization, which complements the equilibrium analysis of the previous section, along with results about the Price of Anarchy. Section~\ref{sec:fluid} deals with fluid approximation models. As opposed to the stochastic queue, explicit results regarding equilibrium and social optimization are often attainable, even for complex systems. Section~\ref{sec:empirical} reports on empirical findings from laboratory experiments that studied decision making related to the timing of arrivals to a queue. A concluding discussion with reference to open problems is given in Section \ref{sec:discussion}.

\noindent{\bf Notation:} Throughout this paper we make use of the following notations: $g'(t)=\frac{\diff}{\diff t}g(t)$ for any differentiable function $g$, $x^+:=\max\{x,0\}$ and $g(t-):=\lim_{s\uparrow t}g(s)$.

\subsection{Related literature}\label{sec:related}

An introduction to basic Queueing Theory that covers the tools required for most of the work described in this survey can be found (among many other books) in Haviv~\cite{book_H2013} or Harchol-Balter~\cite{book_HB2013}, and a more mathematical introduction can be found in Asmussen~\cite{book_A2003}.

The study of strategic decision making in queueing systems goes back to the celebrated Naor model \cite{N1969} who was the first to analyze the decision of whether or not to join a queue from an economic perspective. There are many economic considerations in the analysis of queueing systems, such as joining or balking, purchasing priority, optimal pricing and scheduling and many more. A detailed introduction to strategic queueing can be found in Hassin and Haviv \cite{book_HH2003}. A comprehensive bibliography of this line of research is detailed in Hassin~\cite{book_H2016}, with a special focus on temporal decisions in queueing systems in Chapter 4.1. A related model was studied by Haviv et al. \cite{HKK2010} where customers know their time of arrival and should decide whether to join or not. The analysis requires computing the transient dynamics corresponding to every strategy profile used by other customers, which is also the case in the ?/M/1 model, but ultimately the action of the customer is to either join or not, and this leads to a simpler analysis of the best response function.

The choice of individual arrival time of travelers to a congested bottleneck has been extensively studied in the transportation literature. This goes back to the celebrated Vickrey bottleneck model \cite{V1969} that considers the `rush hour' problem of commuters that need to choose when to start their journey to work while taking into consideration delay, earliness and tardiness costs. This model, as well as most of the subsequent literature, assumes a deterministic fluid model that approximates a system with a large volume of users each having a very small impact on overall congestion. Vickrey's model has been extended and generalized extensively, with Arnott et al. \cite{ADL1993} being a notable example, and a review of this literature can be found in Chapter 11 of \cite{book_DOW2011}. Due to the simpler dynamics, such fluid models have also been frequently used for approximating equilibrium arrival patterns to discrete and stochastic queueing systems. As mentioned above we will discuss the accuracy and applications of such approximations in Section \ref{sec:fluid} of this survey.

\section{Modelling the queue and game}\label{sec:model}

We now elaborate on the general model and highlight some of its important features. We first deal with the queueing process arising from a finite population of customers with independent arrival times. This is followed by discussing the implication of a random population size on the formulation of a non-cooperative game and then a formal definition of a Nash equilibrium for the queueing game. 

\subsection{The queue}\label{sec:queue}

The population size is a random variable $N$ such that $\P(0<N<\infty)=1$, and given $N=n$, the service demands of the $n$ customers are iid random variables $(X_1,\ldots,X_n)$. Conditioning on $N=n$, the arrival times of customers are $n$ independent random variables $(\tau_1\ldots,\tau_n)$ that are possibly not identically distributed. The random arrival time $\tau_i\in\mathcal{T}$, with $\mathcal{T}\subseteq\mathbb{R}$ denoting the admission interval, will later correspond to a strategy played by customer $i$ which can also be represented by the cdf $F_i$. The system processes work during some specified time interval $\mathcal{S}\subseteq\mathbb{R}$. We say that the arrival times are \textit{symmetric} if $F_i=F$ for all $i$, $1 \leq i \leq N$, for some cdf $F$. The above assumptions on arrivals and service hold for all papers discussed in this survey. However, for now we do not assume a specific number of servers or service regime. Of course, the specific system attributes, along with the arrival and service distributions, will determine the distribution of the waiting and departure times (i.e., service completion times) of every customer $i$, $W_i$ and $D_i$, respectively, $1 \leq i \leq N$.

In standard queueing models (without strategic arrival time selection) the arrival process is typically modeled as a renewal process: a sequence $(A_1,A_2,\ldots)$ of iid random variables that represent the inter-arrivals times. For example, the simplest model for arrivals is a Poisson process that assumes $A_i$ are exponential random variables with some fixed rate $\lambda>0$. An accessible introduction to Renewal and Queueing Theory can be found in \cite{book_H2013}. However, when considering a model with a finite population that independently choose arrival times, one must stray from the renewal framework. Suppose that the arrival times of $n$ customers are independent random variables $(\tau_1,\tau_2,\ldots,\tau_n)$, then the inter-arrival times are then given by
\begin{equation}\label{eq:Ai_order}
A_i=\tau_{(i)}-\tau_{(i-1)}, \quad i=1,\ldots, n,
\end{equation}
where $\tau_0:=0$ and $\tau_{(i)}$ denotes the $i$'th order statistic of $(\tau_1,\tau_2,\ldots,\tau_n)$. Clearly the sequence $(A_1,A_2,\ldots)$ is neither independent nor identically distributed. In this case the counting process of arrivals is
\begin{equation}\label{eq:order_statistic_arrivals}
A(t)=\sum_{i=1}^n\mathbf{1}_{\{\tau_{(i)}\leq t\}},
\end{equation}
which is also known as the order statistic process. If $\tau_i$ are identically distributed then the order statistic process is a Markov process (see \cite{C1975}) but not a renewal process as explained above. 

A special case is when the population size~$N$ follows a Poisson distribution with mean $\lambda$. If, in addition, the arrival times
of the participating individuals are independent and uniformly distributed along $[0,T]$, then the arrival process is a Poisson process
with rate $\lambda/T$, which is of course a renewal process. If the population size is Poisson distributed but the arrival distribution is not uniform, but some other symmetric cdf $F_i=F$ for all $i\geq 1$, then the resulting arrival process is a non-homogeneous in time Poisson process as in Example 1. Note that the non-homogeneous Poisson process has independent (but not identically distributed) increments and hence it is not a renewal process. 

A general framework for analyzing non-renewal queueing models with a finite population was introduced by Honnappa et. al \cite{HJW2015}. In particular, they introduced the $\Delta_i$/G/1 model that assumes a deterministic population $N=n$, general service times with CDF $G$, and a single server that operates on a FCFS basis. Using our notation $\Delta_i=A_i$ are the inter-arrival times that are determined by the order statistic process of the arrival times $(\tau_i)_{i=1}^n$ as in \eqref{eq:Ai_order}. A discrete-time version of this queue is analyzed in \cite{M1977}.

For simplicity we now assume that $\mathcal{T}\subseteq\mathcal{S}=[0,\infty)$ and so work is processed as soon as customers start arriving. Therefore, the  workload process satisfies
\begin{equation}\label{eq:workload}
W(t)=\sum_{i=0}^{A(t)}X_i-\int_{0}^t\mathbf{1}_{\{W(u)>0\}}\diff u, \  t \in \mathcal{S},
\end{equation}
and the queue length process satisfies
\begin{equation}\label{eq:queue}
Q(t)=A(t)-S\left(\int_{0}^t\mathbf{1}_{\{Q(u)>0\}}\diff u\right), \ t \in \mathcal{S},
\end{equation}
where $S(t)$ is the uninterrupted service process (counts the number of service completions during an period of continuous work). Note that the  workload represents the waiting time of a potential customer arriving at time $t$ (regardless of if one arrives at this instant or not). If the admission period starts before the service period, then there can be a period when the workload builds up with no work being processed, and then the queueing dynamics in \eqref{eq:workload} and \eqref{eq:queue} need to be  modified accordingly. We will specify these dynamics in the special cases presented in Section \ref{sec:equilibrium}.

The transient analysis of the queue length process $Q(t)$ is generally intractable and so \cite{HJW2015} provides fluid and diffusion limits for the queueing process as $n\to\infty$, with an appropriate scaling of the service time distribution. The limits are obtained by a method of \textit{population acceleration}, which involves increasing the number of arrivals and services during every time interval. For the case of exponential service times, Bet et. al~\cite{BHL2019} show that the queueing process converges in a critically loaded heavy traffic regime to a reflected Brownian motion with non-linear drift. This was later generalized to non-exponential service times by Bet~\cite{B2020}. A similar heavy traffic limit was derived in Bet et. al~\cite{BHL2020} for a queue where the arrival and service times are not independent. Specifically, the arrival times of customers are exponential random variables with a rate that is a linear function of the service time. When the service times are heavy tailed the limiting process was shown to be an $\alpha$-stable process in Bet et. al~\cite{BHL2017}. A \textit{large deviations principle} for extremes of the workload process was established by Honnappa~\cite{H2017}. These limits can provide an approximation for the transient queue dynamics for systems with a large population of customers each having a minuscule  service demand. For example, a sample path version of Little's law for the limiting processes was given in~\cite{B2020}. The fluid limit is especially useful because it enables explicit derivation of the Nash equilibrium arrival strategies in several of the models we will discuss in Section \ref{sec:equilibrium}.

If we make an additional assumption that service times are memoryless; $X_i\sim\exp(\mu)$ for all $i\geq 1$, then the system can be described by a constructing an appropriate continuous-time Markov chain. This construction was first introduced in the ?/M/1 model of~\cite{GH1983} and generalized by Juneja and Shimkin in~\cite{JS2013} and subsequently used in many papers. First assume that the population size is deterministic $N=n$ and that the arrival times $\tau_i$ are continuous and symmetric with cdf $F$, density $f$ and hazard rate $h(t):=\frac{f(t)}{1-F(t)}$. The process $(Q(t),A(t))$, indicating the queue length at time $t$ and the number of arrivals until $t$, is a continuous-time Markov chain. Specifically, this process satisfies the Kolmogorov backward equations
\begin{equation}\label{eq:backward_equations}
\begin{split}
	\frac{\diff}{\diff t}p_{0,j}(t) &= \mu p_{1,j}(t)-(n-j)h(t)p_{0,j}(t), \ 0\leq j\leq n,  \\
	\frac{\diff}{\diff t}p_{i,j}(t) &= \mu p_{i+1,j}(t)+(n-(j-1))h(t)p_{i-1,j-1}(t) -(\mu+(n-j)h(t))p_{i,j}(t), \ 1\leq i\leq j\leq N,
\end{split}
\end{equation}
where $p_{i,j}(t)=\P(Q(t)=i,A(t)=j)$ for $0\leq i\leq j\leq n$, and $p_{i,j}(t)=0$ otherwise. The distributions of the initial conditions $A(0)=Q(0)$ are determined by $F(0)$. This can be generalized to random $N$ by defining the hazard rate after $j$ arrivals as the conditional expectation $h_j(t)=\E[N-j|A(t)=j]\frac{f(t)}{1-F(t)}$. Similarly, the arrival rate can be modified to heterogeneous arrival distribution $F_i$ by conditioning on the subset of customers that have arrived before time $t$.

Further observe that the uninterrupted service process $S(t)$ is now a Poisson process, hence for symmetric arrivals and a deterministic population $n$,  \eqref{eq:queue} yields the expected queue length at time $t$
\begin{equation}\label{eq:E_Qt}
\E[Q(t)]=\E[A(t)]-\mu\int_{0}^t \P(Q(u)>0)\diff u=n F(t)-\mu\int_{0}^t \left(1-\sum_{j=0}^n p_{0,j}(u)\right)\diff u.
\end{equation}
This means that the expected queue length $q(t):=\E[Q(t)]$ and expected  workload 
\[
w(t):=\E[W(t)]=q(t)/\mu,
\]
can be evaluated by solving the backward equations \eqref{eq:backward_equations} given the initial conditions $p_{0,0}(t)=1$.

\subsection{The prior and posterior number of players}\label{sec:prior}

In the examples of Section~1, the symmetric equilibrium was based on the fact that each of the players has the same assumption with respect to number of other players who participate in the game. There is no issue when there is a deterministic number of $N$ players: Each one assumes that the number of other players is $N-1$. But what about the case where this number is random? 

In a model that reflects reality we would like to think of a single lottery which decides on the number of players. Moreover, based on the distribution of this lottery, each of those who in fact were selected by chance to participate, assesses the number of others who were also selected.  We next argue that this number (inclusive of the selected player) needs to be uniquely distributed as the length-biased distribution of the one that governs the lottery. The argument given here is borrowed from Haviv and Oz~\cite{HO2019}. Indeed, suppose there are $m$ potential players. The number of actual  participants is~$i$ with a probability denoted by $\pi_i(m)$, $0 \leq i \leq m$. Assume all $m$ players are equally likely to be selected, so the probability that a tagged player was actually selected, given a total of $i$, is $i/m$, $1 \leq i \leq m$. Hence, by Bayes' rule, the probability that size~$i$ was selected given the tagged one was one of them, equals 
\begin{equation} \label{eq:lbb}
\frac{\frac{i}{m}\pi_i(m)}{\sum_{j=1}^{\infty}\frac{j}{m}\pi_j(m)}, \ 1 \leq i \leq m,	
\end{equation}	
which is easily seen to equal $i\pi_i(m)/\E(N(m))$, where $\E(N(m))$ is the expected number of participants. All one needs to assume is that the limits $\pi_i \equiv \lim_{m \rightarrow \infty}\pi_i(m)$, $i \geq 0$, exist and define a proper distribution. Likewise
with respect to $\E(N)\equiv\lim_{m \rightarrow \infty}\E(N(m)).$ So in the limit, the ratios in~\eqref{eq:lbb} become
\[
\ell_i:=\frac{i\pi_i}{\E(N)}, \ i \geq 1,
\]
which are the length-biased probabilities. In this framework we have the following result for games with a Poisson distributed population (see Haviv and Milchtaich~\cite{HM2012}).
\begin{proposition}\label{prop:poisson_length_bias}
The distribution of the population size $N$ possesses the property that
$\ell_{i+1}=\pi_i$ for all $i\in\{0,1,2,\ldots\}$ if and only if $N$ is a Poisson random variable.
\end{proposition}

Proposition~\ref{prop:poisson_length_bias} states that the Poisson distribution is the unique  distribution over the integers for which the length-biased distribution is the same distribution as the original population size with a shift of one. This extra one is the player who assesses the number of others who participate in the game with one. Therefore, the Poisson assumption simplifies the analysis of games with a random population size, although typically results can be generalized in a straightforward manner by considering the length-biased distribution. Moreover, the Poisson assumption is also attractive because it is a good approximation for the binomial distribution that represents a large pool of customers that potentially join the game, each with a tiny probability. Finally, the Poisson assumption implies the independent increments property of the arrival process, which greatly simplifies the required analysis. Further properties of games with a Poisson distributed number of players are analyzed in the economic literature and we refer interested readers to \cite{MM1987,M1998}.

\subsection{The game}\label{sec:game}

With the queue dynamics and the posterior distribution of the population size at hand, we are ready to define the non-cooperative game and the Nash equilibrium solution for such a game. From now on we assume that $N$ is the  random number of customers joining the queue with respect to the posterior distribution of a customer selected to play as in \eqref{eq:lbb} , i.e., $\P(N=n)=\ell_n$, $n \geq 1$. Any customer~$i$ that has been drawn to play selects an arrival time $t_i \in\mathcal{T}$, $1 \leq i \leq N$. A strategy of customer $i$ is given by a random variable $\tau_i$ with cdf $F_i$, and an arrival profile $\mathcal{F}:=\{F_i,\ i=1,2,\ldots\}$ dictates the strategies of all customers. For any customer $i$ we denote the strategy profiles excluding $i$ by $\mathcal{F}_{-i}$.

From the viewpoint of a tagged customer $i$ the queue length $Q_{-i}(t)$ and his/her waiting time $W_{-i}(t)$ associated with arriving at time $t$, are  functions of the arrival process $A_{-i}(t)$ that is determined by the arrival profile of all other customers $\mathcal{F}_{-i}$. Thus the  expected delay and cost computations are with respect to this process that excludes customer $i$ from the system. If customer $i$ arrives at time~$t$, he/she incurs a cost $C_i(t;\mathcal{F}_{-i},\Theta)$, where $\Theta$ is a set of additional parameters that determine the cost. Denote the expected cost for customer $i$ when arriving at time $t$ by $c_i(t;\mathcal{F}_{-i}):=\E[C_i(t;\mathcal{F}_{-i},\Theta)]$. We next present an example of a cost function for a Markovian queue with penalties on delay and tardiness (based on the results of Haviv \cite{H2013} and Juneja and Shimkin \cite{JS2013}).

\vspace{0.2cm}
\noindent{\bf Example~3.}
A gate to a single server queue is open during the time interval~$[0,T]$, which we call a day. The case $T=\infty$ is not ruled out. Customers are allowed to arrive before time~$0$ and the seniority of those who wish to do so, is kept. This is known as the `with early birds' assumption. They must arrive before time~$T$ in order for their service to be granted (even if it goes beyond time $T$). Therefore, $\mathcal{T}=(-\infty,T]$ and $\mathcal{S}=[0,\infty)$ in the general notation. Service requirements follow an exponential distribution with parameter~$\mu$ and the number of customers who seek service during the day follows a Poisson distribution with parameter~$\lambda$. Customers incur a waiting cost of $\alpha>0$ per unit of time in the queue and a tardiness cost of $\beta>0$ per unit of time from time zero to the time of their service commencement. Hence, the cost function of customer $i$ arriving at $t$ and waiting for $w>0$ units of time equals $\alpha w+\beta\max\{t+w,0\}$. The waiting time consists of the time spent waiting for other customers to be served, and in addition if a customer arrives at $t<0$, before service commences, then he/she has to wait these $-t$ units of time regardless of the queue length at time $t$. Hence, the expected cost is given by
\begin{equation}\label{eq:cost_tardiness}
c_i(t;\mathcal{F}_{-i})=(\alpha+\beta)\E[W_i(t)]+\beta t\mathbf{1}_{\{t\geq 0\}}-\alpha t\mathbf{1}_{\{t< 0\}}.
\end{equation}
If there are $n-1$ other customers with symmetric arrival strategies and service times are exponential with rate $\mu$ then $\E[W_i(t)]=\E[Q_i(t)]/\mu$, where the expected queue length is given by \eqref{eq:E_Qt} (for a system with $n-1$ customers). $\diamond$

We next define a Nash equilibrium, followed by a refinement to a symmetric Nash equilibrium,  with the latter being the main solution concept discussed in this survey.

\begin{definition}\label{def:NE}
A Nash equilibrium is a strategy profile $\mathcal{F}=\{F_i,\ i=1,2,\ldots\}$ such that for any customer $i$ there exists a constant $k_i$ satisfying
\begin{eqnarray*}
c_i(t;\mathcal{F}_{-i}) &=& k_i, \ \forall t\in\sigma(F_i), \\
c_i(t;\mathcal{F}_{-i}) &\geq& k_i, \ \forall t\in\mathcal{T}\backslash \sigma(F_i), 
\end{eqnarray*}
where $\sigma(F_i)\subseteq\mathcal{T}$ is the support of $F_i$.
\end{definition}
Definition \ref{def:NE} is  explained as follows: an individual arrival strategy is optimal in the sense that it is a best response to the strategies selected by the others, and in particular one is indifferent between arriving at all time instants among which the equilibrium strategy is  randomizing.

We next refine the equilibrium definition to a symmetric strategy profile (when possible). Most of the research in the when-to-arrive literature deals with symmetric equilibria, although as we saw in Example~2, there are cases where asymmetric equilibria  exist (even when the game itself is symmetric). Symmetric equilibrium is arguably a more natural outcome for large and anonymous games, and in many cases it is in fact the only possible solution.

\begin{definition}\label{def:NE_symmetric}
A symmetric Nash equilibrium is a strategy $F$ such that there exists a constant $k$ satisfying
\begin{eqnarray*}
c(t;F) &=& k, \ \forall t\in\sigma(F), \\
c(t;F)  &\geq& k, \ \forall t\in\mathcal{T}\backslash \sigma(F), 
\end{eqnarray*}
where $c(t;F)$ denotes the expected cost of any customer who arrives at time~$t$ when all others are using strategy $F$ whose support is $\sigma(F)\subseteq\mathcal{T}$.
\end{definition}

For the sake of brevity, from now on we denote the expected cost by $c_i(t)$, or $c(t)$ in the symmetric case, while keeping in mind that the cost is always a function of the strategy profile. By the above definition we conclude that $\frac{\diff}{\diff t}c(t)=0$ for all $t\in\sigma(F)$ is a condition to be obeyed by  a symmetric and continuous equilibrium arrival distribution $F$.

\section{Equilibrium arrival strategies}\label{sec:equilibrium}

Since the publication of \cite{GH1983} the ?/M/1 model has been generalized and extended in many directions. We first describe in detail the equilibrium analysis of the model with waiting and tardiness costs of Example 3 with the additional assumptions of a Poisson population and exponential service times. The results are due to~\cite{H2013} and~\cite{JS2013}. We find this example a good representative of the formulation and analysis done in many of the subsequent articles extending it or relying on similar analysis. In \cite{JJS2011} and some of the following papers this example is refered to as the `Concert Queueing Game'. Note that the original ?/M/1 model of \cite{GH1983} (and Example 1 here) is a special case with only waiting costs, i.e., $\beta=0$.

\subsection{Equilibrium with waiting and tardiness costs (Example~3)}\label{sec:tardiness}

We recap the model assumptions briefly. A population of customers, whose size $N$ is a Poisson random variable with mean $\lambda$, seek service from a single FCFS server. The admission and service periods are $\mathcal{T}=(-\infty,T]$ and $\mathcal{S}=[0,\infty)$, respectively. We first consider the case `with early birds' where customers are allowed to arrive before time~$0$ while their seniority is kept, and later also review the case `without early birds' where customers are randomly ordered at the opening instant and so there is no incentive to arrive at any time $t<0$. Service requirements follow an exponential distribution with parameter~$\mu$. Customers incur a waiting cost of $\alpha>0$ per unit of time in the queue and a tardiness cost of $\beta>0$ per unit of time from time zero until their service commencement. 

Customers decide on their arrival time, $t$, which needs to be in the time interval $[-t_a,t_b]$ for some (to be determined) $t_a \geq 0$ and $0 \leq t_b \leq T$. Their goal is to minimize their expected total (waiting plus tardiness) costs given in \eqref{eq:cost_tardiness}. Since mixing is allowed (and in fact, needed) we seek a symmetric arrival strategy which states a CDF $F(t)$ that constitutes a Nash equilibrium. Given a symmetric  arrival strategy $F$, denote the expected queueing time for one who arrives at time $t$ by $w(t)$. Then the expected cost associated with arriving at time $t$ is given by \eqref{eq:cost_tardiness},
\[
c(t)=(\alpha+\beta)w(t)+\beta t\mathbf{1}_{\{t\geq 0\}}-\alpha t\mathbf{1}_{\{t< 0\}},
\]
where $w(t)=\E[Q(t)]/\mu$. We make the following observations on the equilibrium distribution:
\begin{itemize}
\item {\bf No atoms:} The equilibrium strategy $F$ does not come with atoms. This is because arriving momentarily before an atom ensures a lower expected queue length, i.e., if $F(t)$ has an upward discontinuity at $t$ then so do $\E[Q(t)]$ and $w(t)$. Hence, we look for a CDF which comes with a density function $f(t)$ such that $F(t)=\int_{\tau=-t_a}^t f(\tau)\,d\tau$, where $f(\tau)$ is continuous almost everywhere. 
\item {\bf No holes:} There are no gaps in the support of the arrival time distribution: If a gap exists, one better arrive at its end rather than at it initiation. This is verified by taking derivative of the cost function $c(t)$ (see Lemma 2.1 of \cite{H2013} for details).
\end{itemize}

Given the above properties, Definition \ref{def:NE_symmetric} states that in equilibrium there exists some constant $k>0$,
\begin{equation}\label{eq:exmp3_equil}
(\alpha+\beta)w(t)+\beta t=k, \ \ t\in[-t_a, t_b],
\end{equation}
and
\begin{equation} \label{eq:exmp3_equil2}
(\alpha+\beta)w(t)+\beta t \geq k, \ \ t\notin[-t_a, t_b].
\end{equation}
The first thing to observe is that $k=\alpha t_a$ as this is the cost of one who arrives at time $-t_a$ and faces an empty queue with probability one. Since,
\[
w(t)=\frac{\E[Q(t)]}{\mu}=\frac{\lambda F(t)}{\mu}-t, \ \ -t_a \leq t < 0,
\]
we have that
$$\a\frac{\lambda F(t)}{\mu}-t+\b \frac{\lambda F(t)}{\mu}$$
is constant along the time interval $[-t_a,0]$.
Hence, 

\[
f(t)=\frac{\mu}{\lambda}\frac{\alpha}{\alpha+\beta}, \ \ -t_a \leq t \leq 0.
\]
In particular, the distribution along the early birds arrival period is uniform.
Therefore, $F(0)=t_a \mu \alpha/(\lambda(\alpha+\beta))$ and
\begin{equation} \label{eq:marg}
\int_{0}^{t_b}f(t)\,dt=1-t_a \frac{\mu}{\lambda}\frac{\alpha}{\alpha+\beta}. 
\end{equation}
Assume for now that $T=\infty$. This assumption will be removed shortly.

Clearly, the density function $f(t)$ determines the arrival process, which due to the Poisson assumption regarding the number of arrivals during the day, is a  non-homogeneous Poisson process with a rate function $\l f(t)$. Moreover, the
future progression of the number who queue up at time $t$, which in turn determines the expected waiting time given an arrival at this instant, does not depend on the past, once the current number in the system is given. Specifically, denote by $p_i(t)$, the probability that at time~$t$, $t \geq 0$, the number in the system is~$i$, $i \geq 0$. Then, these probabilities obey the initial conditions, given by the splitting property of the Poisson distribution,
\begin{equation} \label{eq:inicon}
p_k(0)=e^{-\lambda F(0)}\frac{(\lambda F(0))^k}{k!}, \ \  k \geq 0
\end{equation}
and the dynamics described in \eqref{eq:backward_equations} are
\begin{eqnarray}
p'_0(t) &=& p_1(t)\m - p_0(t)\l f(t), \ \ 0 < t < t_b ,  \label{eq:dynamics1} \\
p'_k(t) &=& p_{k-1}(t)\l f(t) + p_{k+1}(t)\m - p_k(t)(\l f(t)+\m), \ \ 0 <t < t_b, \ k \geq 1. \label{eq:dynamics2}
\end{eqnarray}

The standard queue dynamics described in \eqref{eq:E_Qt} yield
\[
w'(t)=\frac{1}{\m}(\l f(t)-\m(1-p_0(t))).
\]
This coupled with the equilibrium condition~(\ref{eq:exmp3_equil}) turn out to be equivalent to
\begin{equation} \label{eq:exmp3_f_equil}
\begin{split}
f(t)=\left\lbrace\begin{array}{ll}
\frac{\mu}{\lambda}\frac{\alpha}{\alpha+\beta},&  -t_a \leq t < 0,  \\
\frac{(1-p_0(t))\m}{\l}-\frac{\b \m}{(\a +\b)\l},&  0 \leq t \leq t_b .
\end{array}\right.
\end{split}
\end{equation}
Another equilibrium condition is that 
\begin{equation} \label{eq:exmp3_equil3}
f(t_b)=0,
\end{equation}
as without it one better arrive at time $t_b+\epsilon$ for some small $\epsilon>0$, violating~(\ref{eq:exmp3_equil2}). 

The above provides a full characterization of a symmetric equilibrium arrival distribution but the complicated queueing dynamics do not enable an explicit solution of the functional differential equations. We next describe a simple method for numerical computation of the equilibrium distribution.

\begin{algorithm}[H]
\renewcommand{\thealgorithm}{}
\caption{\textbf{1}: Numerical computation of the equilibrium strategy $F$.}
\begin{algorithmic}\label{algo:F_NE}
\State Set $k=0$ and guess $t_a^{(k)}>0$.
\vspace{0.1cm}
	\State \ \ (a) Compute $f(t)$ for $t<0$ and $F(0)$ from \eqref{eq:marg} and then $p_k(0)$ using (\ref{eq:inicon}).
	\State \ \ (b) Compute $f(t)$ for $t>0$ using \eqref{eq:exmp3_f_equil}, \eqref{eq:dynamics1} and \eqref{eq:dynamics2}, stopping when $t_b$ satisfies \eqref{eq:marg}.
	
	\vspace{0.1cm}
\State If $f(t_b)>0$ then set $t_a^{(k+1)}>t_a^{(k)}$ and repeat steps (a) and (b).
\State If $f(t_b)<0$ then set $t_a^{(k+1)}<t_a^{(k)}$ and repeat steps (a) and (b).
\State If $f(t_b)=0$ then \eqref{eq:exmp3_equil3} is satisfied and $F$ is the equilibrium distribution.
\end{algorithmic}
\end{algorithm}

Of course, implementation of Algorithm 1 requires some more details. In practice, it is convenient to do the computation on a discretized time grid and to use a smart bisection search for finding $t_a$ quickly. The converegence of the algorithm to a unique solution is guaranteed by monotonicity properties discussed in \cite{H2013,JS2013}. It is further established in \cite{JS2013} that the symmetric equilibrium solution here is the unique solution, even if asymmetric strategies are allowed. The uniqueness result requires a technical assumption that mixed strategies with an infinite number of atoms in a finite interval are not allowed.

One more feature the of equilibrium density is that it comes with a downwards discontinuity at $t=0$; $f(0-)>f(0)$. This phenomenon is repeated in other when-to-arrive models. It is interesting that the equilibrium arrival rate drops as soon as the server commences service.

\begin{figure}[]

\begin{subfigure}{.98\linewidth}
\centering
\begin{tikzpicture}[xscale=0.4,yscale=30]
  \def\xmin{-15}
  \def\xmax{15}
  \def\ymin{0}
  \def\ymax{0.1}
    \draw[->] (\xmin,\ymin) -- (\xmax,\ymin) node[right] {$t$} ;
    \draw[->] (0,\ymin) -- (0,\ymax) node[above] {$f(t)$} ;
    \foreach \x in {0}
    \node at (\x,\ymin) [below] {\x};
    \draw[blue,  thick, domain=\xmin:-10] plot (\x, {0});
	\draw[blue,  thick, domain=-10:0] plot (\x, {0.06});
	\draw[blue,  thick] (0, 0.057) -- (0.10359,0.05696) -- (1.13949,0.05693095) -- (2.29049,0.05598307) -- (3.44149,0.05404521) -- (4.59249,0.05107672) -- (5.74349,0.04704752) -- (6.89449,0.04182389) -- ( 8.04549,0.03508859) -- (9.19649,0.02621195) -- (10.34749,0.01394281)  -- (11.28,0) ;	
	\draw[blue,  thick, domain=11.26829:\xmax] plot (\x,{0});
	\draw[blue,-, densely dashed] (0,0.06) -- (0, 0.057);
	\draw[blue,-, densely dashed] (-10,0) -- (-10, 0.06);
	
	\node[below] at (-10,0)  {\textcolor{blue}{\small{$t_a=-10$}}};
	\node[below] at (11.2,0)  {\textcolor{blue}{\small{$t_b=11.2$}}};
\end{tikzpicture}
\end{subfigure}

\begin{subfigure}{.98\linewidth}
\centering
\begin{tikzpicture}[xscale=0.4,yscale=3]
  \def\xmin{-15}
  \def\xmax{15}
  \def\ymin{0}
  \def\ymax{1.05}
    \draw[->] (\xmin,\ymin) -- (\xmax,\ymin) node[right] {$t$} ;
    \draw[->] (0,\ymin) -- (0,\ymax) node[above] {$F(t)$} ;
    \foreach \x in {,0}
    \node at (\x,\ymin) [below] {\x};
    \foreach \y in {1}
    \node at (0,\y) [below left] {\y};
    \draw[blue,  thick, domain=\xmin:-10] plot (\x, {0});
	\draw[blue,  thick] (-10,0) -- (0,0.58949);
	\draw[blue,  thick] (0,0.58949) -- (0.10359,0.5952034) -- (1.13949,0.6465591) -- (2.29049,0.7030807) -- (3.44149,0.7581633) -- (4.59249,0.8107884) -- ( 5.74349,0.8599155) -- (6.89449,0.9044282)  -- ( 8.04549,0.9429915) -- (9.19649,0.9738104) -- (10.34749,0.9941864) -- (11.28,1) ;	
	\draw[blue,  thick, domain=11.26829:\xmax] plot (\x,{1});
	
	\draw[black,-,  dashed] (\xmin,1) -- (\xmax, 1);
	\draw[black,-,  dashed] (11.2,0) -- (11.2, 1);
	
	\node[below] at (-10,0)  {\textcolor{blue}{\small{$t_a=-10$}}};
	\node[below] at (11.2,0)  {\textcolor{blue}{\small{$t_b=11.2$}}};
\end{tikzpicture}
\end{subfigure}
\caption{Example 3 (waiting and tardiness costs). The cdf and density of the Nash equilibrium arrival distribution. The example parameters: $\left(\lambda=20, \ \mu=2, \ \alpha=1.2, \ \beta=0.9\right)$.}  \label{fig:F_NE}
\end{figure}
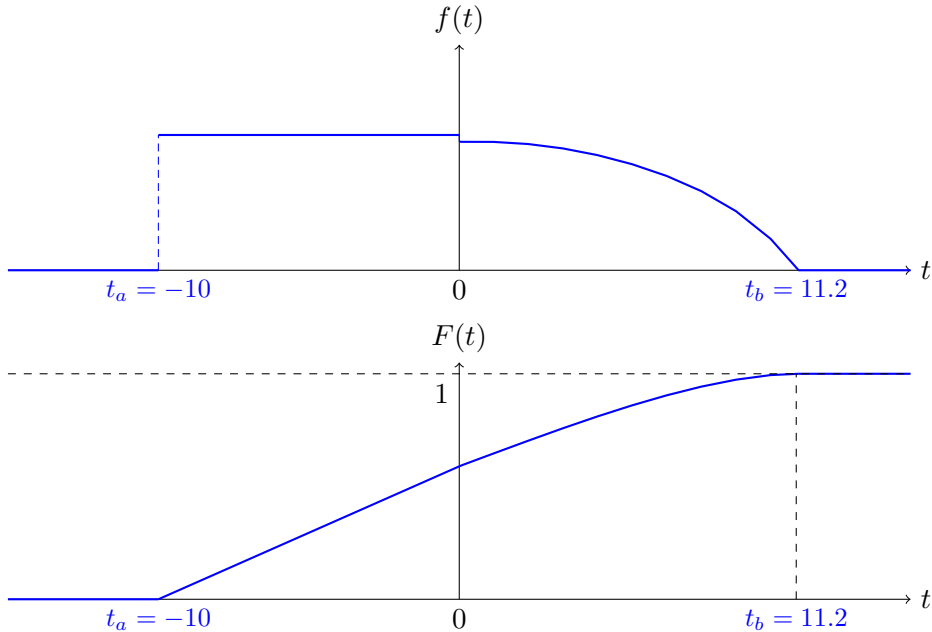

Figure \ref{fig:F_NE} illustrates the density and the cdf which determine the equilibrium arrival process. In this example the parameters are $\left(\lambda=20, \ \mu=2, \ \alpha=1.2, \ \beta=0.9\right)$. Customers start arriving at $t_a=-10$ and they arrive before the opening with a probability of $F(0)=0.59$. After the opening customers arrive with a decreasing to zero density until $t_b=11.2$. Recall that the system may continue working after $t_b$ if there are still customers waiting for or at service at time $t_b$.

A final remark for this example is that the analysis straightforwardly  extends to any distribution on the population size,
$N$. As explained in Section \ref{sec:game}, the dynamic equations \eqref{eq:dynamics1} need to be replaced by \eqref{eq:backward_equations} for the two-dimensional process $(A(t),Q(t))$ representing the number of arrivals and the queue length, respectively. In particular, the arrival rate at time $t$ when $A(t)=j$ customers have already arrived is given by the hazard rate function
\[
h_j(t)=\E[N-j|A(t)=j]\frac{f(t)}{1-F(t)}.
\] 
For any $N=n$ the number of customers in the queue at $t=0$ is a Binomial$(n,F(0))$ random variable, hence the initial conditions of the system are now given by 
\[
p_k(0)=\sum_{n=1}^\infty \ell_n {n \choose k} F(0)^k (1-F(0))^{n-k}, \ k\geq 0,
\]
where $\ell_n$ is the length-biased distribution defined in Section~\ref{sec:prior}. If $N$ is a constant then $\ell_{N-1}=1$ for any tagged customer.

\subsection{Restrictions on arrival times}\label{sec:arrival_constraints}

In Example~1, where only waiting costs ($\beta=0$) were assumed, it was necessary to assume a finite closing time $T<\infty$; otherwise, customers could `spread'  their distribution over an unbounded period leading to virtually zero interaction and no waiting costs. When tardiness costs exist as well, then even without a restriction on the  customers to arrive on a finite interval, the resulting arrival period will be bounded. This is because the tardiness cost is unbounded for arriving very late. However, one may still want to consider a system with a closing time and tardiness costs and the analysis in Section \ref{sec:tardiness} still holds with a slight modification. It is clear that if $T<\infty$ but still $T>t_b$, then the equilibrium distribution remains the same. Otherwise, $t_b$ needs to be replaced by $T$ throughout the analysis (removing one degree of freedom from the solution) and the equilibrium condition~(\ref{eq:exmp3_equil3}) needs to be removed. For further details on the construction of the equilibrium in this case we refer to \cite{H2013}.

Suppose now that early birds are not allowed and all customers present at the opening instant are randomly ordered. This may be the case in a facility without a physical space to form a queue before the gates are open, or in digital systems that are simply turned on at a certain moment in time. The game with no early birds was first studied by Hassin and Kleiner \cite{HK2011} for a model with only waiting costs and later extended to a model with waiting and tardiness costs in \cite{H2013}. In this case there is no reason for a customer to arrive at any $t<0$ and the equilibrium distribution is therefore restricted to $[0,T]$. The key difference in the solution is that now there is an atom at time zero whose size is $F(0)>0$. Furthermore, as there is an upward discontinuity in the waiting time immediately after the atom, there will be no arrivals  along a positive interval immediately after the opening. Formally, there exists some $t_e>0$ such that $F(t_e)=F(0)$ and $\int_{t=t_e}^T f(t)\diff t=1-F(0)$. Note that it is possible that $t_e=T$, i.e., all customers arrive together at the opening with probability one (as was shown in Example~2 for a game with two customers). The density $f(t)$ on $[t_e,T]$ is obtained in the same manner as was exemplified in the case without early birds, with initial conditions that are given by \eqref{eq:inicon}, with $F(0)$ as the boundary parameter instead of $t_a$. Therefore, Algorithm~1 can still be applied with little modification.

\subsection{Computing the equilibrium}\label{sec:comp}

Explicit analysis and computation of the equilibrium distribution is impossible for most models and so algorithmic and numerical approaches are often used. The most commonly used  method for computing the equilibrium is via a discrete approximation of the solution to the functional differential equations arising from the equilibrium conditions together with the queueing dynamics. We provide details on this approximation for the model with waiting and tardiness costs of Example~3. 

Suppose that early birds are not allowed and that the system has a finite closing time, i.e., $\mathcal{T}=[0,T]$. One can choose a small $\Delta$ such that $T/\Delta$ is an integer and solve the game on a discrete grid $\{0,\Delta,2\Delta,\ldots,T\}$. The population size, and therefore also the queue, is unbounded due to the infinite support of the Poisson distribution and so a truncation of the state space is in order. Namely, assume that the queue buffer is a large integer, denoted by~$M$. The queue dynamics of \eqref{eq:dynamics1} and \eqref{eq:dynamics2} are then approximated for every $t\in\{0,\Delta,2\Delta,\ldots,T\}$ by
\[
p_0(t+\Delta) = p_0(t)+\Delta (p_1(t)\mu - p_0(t)\lambda f(t))
\]
and
\[
p_k(t+\Delta) = p_k(\Delta)+\Delta( p_{k-1}(t)\lambda f(t) + p_{k+1}(t)\mu - p_k(t)(\lambda f(t)+\mu)),\  1\leq k \leq M,
\]
where $p_{M+1}(t):=0$ for any $t$. This approximation is justified by recalling that one of the definitions of a continuous-time Markov chain is that transitions from state $i$ occur at time $t$ with probability of the order $\Delta q_i(t)+o(\Delta)$, where $q_i(t)$ is the total transition rate from state $i$ at time $t$. 

In Algorithm~1, for any initial value $F(0)$, the probabilities $p_k(t)$ need to be evaluated for all $k\in\{0,\ldots,M\}$ and $t\in\{0,\ldots,T\}$. We therefore conclude that a single iteration of Algorithm~1 requires an order of $M\cdot T/\Delta$ computations. For an arbitrary initial guess of $F(0)$ multiple iterations are typically required until convergence is reached, but typically the number of iterations is small and the time consuming part of the algorithm are the iterations themselves. Increasing $M$ or decreasing $\Delta$ clearly improve the accuracy of the approximation but both come at the cost of increasing the computation time of every iteration of the algorithm.

\subsection{Cost functions}\label{sec:cost}

The existence and uniqueness of the symmetric Nash equilibrium is rigorously verified in \cite{JS2013} for the game with early birds. As illustrated in Example 2, if early birds are not allowed and $T<\infty$ there may exist multiple asymmetric equilibria where some customers arrive at time zero and some at $T$. Excluding such cases, the underlying result is that the equilibrium dynamics given by demanding the cost $c(t)$ to be constant along the support and the backward equations of the queueing processes yield a unique solution that is monotone with respect to initial conditions. 

Breinbjerg \cite{B2017} extends the existence and uniqueness result to a model with general (non exponential) service times and a non-linear cost function. Specifically, a utility function that is continuous and decreasing with respect to both waiting and departure times are assumed. A constructive characterization of the symmetric equilibrium was provided and uniqueness was argued by applying the aforementioned monotonicity properties of the equilibrium dynamics. The construction of the equilibrium for general service times and non-linear utilities does not provide a general tractable computational method, so specially tailored numerical methods for the specific dynamics and cost structure need to be constructed for every model. An algorithm for computing the Nash equilibrium in a discrete-time setting with general service times is constructed by Sakuma et. al \cite{SMF2019} for the model with no early birds and $\beta=0$.

In Ravner \cite{R2014} it is assumed that the customers incur a cost based on the order of admission. This is the case for example in a concert hall or a bus with unassigned seats. Let $\gamma>0$ denote the linear order penalty, i.e., a customer arriving at time $t$ incurs a cost of $\gamma A(t)$, where $A(t)$ is the number to appear by time $t$. Suppose that their additional tardiness and waiting costs are as before, then the equilibrium has the same structure as in Example~3, with a modification of the density function \eqref{eq:exmp3_f_equil} that takes into account the additional cost parameter $\gamma$. The case of $T=\infty$ and $\beta=0$ is interesting because the equilibrium distribution has an infinite support with a density that decreases exponentially fast to zero but never equals zero. Therefore, in this case the arrival process can last for a very long time (with a positive probability): Even if one arrives late (long after all other customers have arrived), the expected cost is approximately $\gamma N$ because the queue is most likely empty. This further implies that in equilibrium the expected cost for all customers is exactly $\gamma N$. For the special case of $N=2$ the equilibrium distribution has an explicit solution which turns out to be an exponential distribution after time zero.

Some service systems are always operational and don't have an opening time, but may still be congested for only a specific time period. For example, the highway leading to a city is always open but traffic jams occur mostly at rush hour when the demand for getting into the city is highest. This is also the case for various digital systems such as streaming services with peak demand at certain hours. A natural model for such a system is presented by Sherzer and Kerner in \cite{SK2017}: a finite population of $N$ customers that can arrive any time $t\in(-\infty,\infty)$, but now there is also a penalty for arriving early. The costs are linear with $\beta_1$, $\beta_2$ and $\alpha$ denoting the per unit of time earliness, tardiness and waiting costs, respectively. The ideal time for arriving and starting service is assumed to be $t=0$, without loss of generality. The expected total costs for a tagged customer arriving at time $t$ is then
\begin{equation}\label{eq:cost_early}
c(t)=\left\lbrace\begin{array}{ll}
-t\beta_1+\alpha w(t)+\beta_2\E\left[(t+\sum_{i=1}^{A(t)}X_i)^+\right], & t<0, \\
(\alpha+\beta_2) w(t)+\beta_2 t, & t\geq 0,
\end{array}\right.
\end{equation}
recalling that $w(t)$ is the expected waiting time of one who arrives at time $t$ and $X_i$ is the service time of  customers~$i$. Note that earliness is the (negative) deviation of the arrival time from zero and tardiness is the deviation of the time of starting service from zero and therefore a customer may arrive early and still incur a tardiness cost if their waiting time ends after zero. For a Poisson population, using similar arguments as in Example 3, \cite{SK2017} provides a similar characterization of a unique symmetric arrival distribution: $F$ is continuous with a density that satisfies a system of differential equations given by the queue dynamics and the constant expected cost equilibrium condition. However, the equilibrium solution is harder to compute because the density before zero is no longer uniform and the derivative of \eqref{eq:cost_early} is a function of the probabilities $p_k(t)$ of all states $k\geq 0$ and not just the idle probability $p_0(t)$ as in \eqref{eq:exmp3_f_equil}. The solution can be approximated with a procedure similar to Algorithm~1.

An additional paper that assumes earliness costs is Glazer et. al \cite{GHR2017} for a machine scheduling motivated model. A single machine processes $N$ determinsticly-sized jobs that have a desired due date with penalties for tardiness and earliness, but no cost for waiting. It is shown that multiple equilibria exist such that all jobs arrive at the same time. For most parameter settings the socially optimal schedule is also an equilibrium. The above result holds for several extensions such as heterogeneous jobs that may differ in their due dates or earliness/tardiness penalties.

\subsection{Service regimes and queue dynamics}\label{sec:other_dynamics}

Timing of arrivals is also relevant for periodic batch-service systems, such as a bus or train that departs at some known time. Such a model is studied in Glazer and Hassin \cite{GH1987}: service is provided to a batch of up to $N\geq  1$ customers every $T>0$ units of time, and if there are more than $N$ customers waiting at the time of service then only the first $N$ are served (with FCFS order) and the remaining customers must wait for at least another $T$ units of time until the next service. Every period a Poisson number of customers joins the system and choose when to arrive during $[0,T]$. Note that the batch sizes are independent but the waiting times are not as at the beginning of a period there may be customers left from previous periods. The steady-state arrival distribution is shown to be a continuous distribution $F$ on $[t_e,T]$ for some $t_e>0$ with a density $f(t)$ that satisfies a system of backward equations, similar to those detailed for Example~3 above. 

Lariviere and Van Mieghem \cite{LvM2004}  consider a  a discrete-time system with $M$ customers that choose between $T$ possible arrival slots. Each time slot is assumed to be sufficient to serve all arriving customers so there are no queue dynamics and the expected congestion cost is based only on the number of other customers with the same chosen time slot. The arrival process corresponding to the mixed Nash equilibrium is shown  when  $M$ and $T$ grow large, to be approximately a homogeneous in time Poisson process.  

Suppose now that the system has a finite buffer and customers arriving when the buffer is full are lost and do not obtain service. In Mazalov and Chuiko \cite{MC2006} a single server with exponentially distributed service times and no waiting buffer is considered, i.e., an ?/M/1/1. The utility of a customer arriving and finding the sever idle at time $t$ is given by some continuous function $C(t)$, and the utility is zero if the customer finds the server  busy and the customer is then  lost. The mixed strategy Nash equilibrium is again given in the form of a continuous distribution that solves a system of differential equations with some special cases admitting a closed form solution. Haviv and Ravner \cite{HR2015} assume that customers wish to maximize the probability of admission in a multi-server finite buffer queue (an ?/M/$m/c$ system) by choosing an arrival time $t\in[0,T]$. As in the no early birds case of Section \ref{sec:arrival_constraints}, the symmetric equilibrium strategy has an atom at zero $F(0)$ and a continuous distribution on an interval $[t_e,T]$. If there is a single server and no waiting buffer; $m=c=1$, then the equilibrium distribution is uniform for the special cases of $N=2$ and $N$ following a Poisson distribution. Note that for $N=2$ this is exactly Example~2. This is the case  as there is only one other potential customer, we get that minimizing  waiting is equivalent to maximizing the probability  of finding an idle server. The model of \cite{HR2015} is extended by Chuiko~\cite{C2017} to a loss system with two servers and random routing. A similar equilibrium distribution is derived by solving a more elaborate system of equations that takes into account the transient joint distribution of the state of both servers.

So far we have discussed systems with FCFS policy. However, there are  many systems that employ other service regimes such as priority classes, processor sharing or last come first served (LCFS). Such a policy may be in order for a system with heterogeneous customers or as a means for improving performance. The queue dynamics become much more elaborate for non-FCFS regimes, for example overtaking implies that the expected waiting time of a customer arriving at $t$ depends on arrivals after time $t$ and so a recursive solution of the equilibrium dynamics as in Algorithm~1 no longer works. Therefore, approximations of the equilibrium or of the system dynamics are often used for studying such models. The most common approach is by applying fluid approximations. This line of work will be discussed in detail in Section~\ref{sec:fluid}.

A classical result of Hassin \cite{H1985} states that LCFS preemptive-resume policy is socially optimal for an observable queue with customers that can balk from the queue. Similar results  are obtained for the game of timing arrivals by Platz and {\O}sterdal~\cite{PO2017} and Breinbjerg and {\O}sterdal~\cite{BO2017}. In particular, \cite{PO2017} establishes the optimality of the LCFS, and, interestingly, that FCFS is the worst in a wide class of service regimes. This is done by considering a fluid approximation of the game. \cite{BO2017} constructs a numerical method to compute the equilibrium arrival distribution for LCFS. It is further shown numerically that the social welfare in equilibrium is higher for LCFS than FCFS. An additional model that is related to LCFS is the timeline game of Altman and Shimkin \cite{AS2016d}. This model considers a website with ad listings, such as real-estate properties, where the newest ad bumps down all existing ads. The game takes place over a finite duration $[0,T]$ and advertisers choose when to upload their ad with the  goal that their ad be placed as high as possible for as long as possible. It is shown there that the symmetric equilibrium distribution is given by a continuous distribution on $[0,t_e]$ for some $t_e<T$ and is derived by using similar equilibrium conditions as the ?/M/1 model. 

An observable version of the when-to-arrive game is studied in Babaioff and Oren~\cite{BO2018}. A population of $N$ customers choose when to arrive from a discrete set of times, $\mathcal{T}=\{1,2,\ldots\}$, and customers arriving at the same time are randomly ordered. Before every unit of time all customers who have not joined can observe the queue and decide whether to join or wait. The service time is deterministic and equals one unit of time for each customer. Customers wish to minimize their waiting time in the queue. This setup yields a stochastic game and therefore the solution concept analyzed is the subgame perfect equilibrium (SPE). It is shown by induction that a symmetric SPE always exists. This is done by the construction of the discrete queue dynamics and the evolution of the expected costs. If the waiting cost is low, specifically a linear cost of $\alpha \leq 2$ per unit of time, then the equilibrium prescribes that all customers join in the first period with probability $1$. For $\alpha>2$ there exist more elaborate mixed strategies that randomize between joining or not for several time periods. Of course, every such strategy dictates what action to make for every queue length observed and therefore the equilibrium strategies can be described algorithmically but not explicitly.

A model allowing for heterogeneous customers and priorities was introduced by Talak et. al \cite{TMP2019}. The paper studies the problem of when to arrive at the gate for boarding a plane. The gate closes at $t=0$ and so all customers must choose an arrival time $t\leq 0$ and wait behind all travelers who have already joined the queue, i.e., $\mathcal{T}=(-\infty,0]$ and $\mathcal{S}=\{0\}$. It is assumed that the cost of a customer is a weighted combination of the queueing time and penalty for arriving at the gate $-t$ units before boarding commences. The weight of the waiting time penalty, denoted by $n_v$, is determined by the type $v$, which is given by a continuous random variable from a known common distribution. Customers decide when to arrive given their individual waiting cost. The unique Nash equilibrium dictates a deterministic arrival time for each waiting penalty realization. The paper considers several extensions and generalizations. A notable extension allows customers to purchase priority, from a discrete set of $L$ priority levels, as well as choosing when to arrive. Under some assumptions it is shown that, in equilibrium, each priority level $\ell=1,\ldots,L$  comes with an associated interval of customer types: every customer of type $v\in [v_{\ell},v_{\ell}]$ purchases priority $\ell$, and furthermore the customers at each level  have disjoint arrival intervals. Moreover, customers with a higher weight $n_v$ for the waiting penalty arrive later and purchase higher priorities than those who have a lower weight.

In many queueing systems, such as those arising in traffic and communication networks, the service regime is processor sharing rather than FCFS. As explained before, solving the equilibrium dynamics for such a system is much harder even for a Markovian system because the waiting time of a customer arriving at time $t$ is determined by the number of customers in the system at $t$ and all those arriving after time $t$ (and before the customer departs). A fluid approximation for this model will be discussed in Section \ref{sec:fluid}. In Ravner et. al~\cite{RHV2016}, a Stackelberg game of choosing arrival times to a deterministic processor sharing system is analyzed. Every customer has a unit job size and an individual due date, and the cost is a weighted combination of the sojourn time in the system and the deviation from the due date. The customers choose the arrival times sequentially in the order of their due dates. It is shown that multiple subgame perfect equilibria exist for this game. Furthermore, an iterative best-response algorithm for computing equilibrium points is constructed. 

The aforementioned \cite{SMF2019} considers a discrete-time version of \cite{HK2011} (a FCFS system with no early birds and penalty only for waiting) with general service times. The equilibrium distribution are numerically compared to the outcome of a dynamic agent based learning model: A large population of potential customers join the queue every day independently, each with some small probability. If they indeed join, they need to choose when to arrive. Customers start by choosing an arrival uniformly on the first day and record the waiting time experienced at the chosen arrival time. As time evolves customer mix between arriving uniformly and choosing the time that yielded the minimal average waiting time in the past, with a growing probability for the latter. The above exploration-exploitation dynamics converge to an empirical distribution that is shown by means of simulation to be very close to the Nash equilibrium arrival distribution. Intuitively, all the time slots that are chosen repeatedly in the long run  yield the same waiting time and therefore this approximates the equilibrium condition of a constant expected cost over the support of the distribution. This type of learning dynamics can be a powerful tool to approximate the equilibrium arrival distribution for more complicated systems which are otherwise intractable. Ravner and Sakuma \cite{RS2020}, extend this learning model to a system with a random service rate and noisy signals. In particular, customers are heterogeneous due to receiving different information about the service speed and computing the equilibrium becomes a difficult task even for a simple model with two types of customers and exponential service times. The long term outcome of the learning dynamics are shown to be quite differ from the equilibrium outcome when customers have limited information about the system parameters.

\section{Social optimality and Price of Anarchy}\label{sec:optim} 

So far we have formulated the ``when-to-arrive" problem as a non-cooperative game and the solution concept we looked for is that on symmetric Nash equilibrium. Now we consider a different criterion, that of society. As elaborated in Section~\ref{sec:model}, the strategy profile $\mathbf{T}=\{\tau_1,\ldots,\tau_n\}$ determines  the random arrival process to the system. Given the arrival process every customer $i=1,\ldots,n$ has some expected cost $c_i(\mathbf{T})$ and the social cost is defined as the sum of the individual costs, $\sum_{i=1}^n c_i(\mathbf{T})$. Consider some symmetric (or for that matter, asymmetric) strategy profile that associates a cost for each of the players. A socially optimal arrival profile is then
\[
\mathbf{T}^*\in\argmin_{\mathbf{T}} \sum_{i=1}^n c_i(\mathbf{T}).
\] 
It is important to make a distinction between two levels of control the optimizer has over the arrival times:
\begin{itemize}
\item {\bf Full control} The individual arrival times $(t_1,\ldots,t_n)\in\mathbb{R}^N$ of all customers can be selected by a central planner. This means that all customers can be identified and instructed when to arrive. In particular, symmetric customers can be told to arrive at different times. If the population size $N$ is random then its realization should also be known.
\item {\bf Optimal symmetric strategy} A single message can be sent to all customers instructing them what, possibly mixed, strategy to use. This is an anonymous control. If customers are heterogeneous in the sense that they  belong to different classes,  the message may specify a different strategy for every type of customer. An assumption invoked here  is that customers know their own type.
\end{itemize}

The optimal full control schedule of arrivals can be seen as an upper bound on the social welfare because no static policy can do better. The anonymous control, on the other hand, is limited to solutions that can be implemented by the customers themselves without centralized coordination. Customers are still anonymous but instead of a Nash equilibrium profile, the symmetric individual strategy is chosen in such a way that social welfare is maximized. Therefore, the latter is a more natural benchmark for the equilibrium social welfare because it maintains the rules of the game.
 
The inefficiency of the system is typically measured by the Price of Anarchy (PoA) , which is the ratio between the worst equilibrium cost and the socially optimal cost,
\[\mathrm{PoA}=\frac{\sup_{\mathbf{T}^e\in\mathcal{T}^e}\sum_{i=1}^n c_i(\mathbf{T^e})}{\sum_{i=1}^n c_i(\mathbf{T^*})}\geq 1,
\]
where $\mathcal{T}^e$ is the set of all Nash equilibria. In most cases $\mathbf{T}^*$ is the optimal symmetric control, but in some cases the full control optimal solution is easier to compute and then taking $T^*$ as this solution can be viewed as an upper bound on the PoA.

\noindent{\bf Example~2 (cont.)}
Recall that two customers seek service from a single service and choose when to arrive during the time interval $[0,T]$. Service times are exponential with rate $\lambda$. We next detail the equilibrium, full control and anonymous control solutions.
\begin{itemize}
\item {\bf Symmetric equilibrium} If $T<\log 2/\mu$ then both customers arrive at $t=0$. If $T>\log 2/\mu$ then the symmetric equilibrium strategy is given by $F^e(0)=2/(2+\mu T-\log 2)$ and a uniform density of $\m/(2+\m T - \log 2)$ along the interval $[\log 2/\mu,T]$. Note that the above ignores the possible asymmetric equilibria discussed in Section \ref{sec:intro}. If both customers arrive at $0$ the social cost is $1/\mu$ and for the symmetric mixed equilibrium it is $4/\mu(2+\mu T-\log 2)$.
\item {\bf Optimal Symmetric strategy} If we limit the profiles to symmetric ones, it is shown by Haviv and Oz~\cite{HO2018} that the (practically) unique optimal solution is to arrive with probability 
\[
F^*(0)=F^*(T)-F^*(T-)=1/(2+\mu T)
\]
at either time~$0$ or time~$T$, and with a uniform density of $\mu/(\mu T+2)$ along the interval~$(0,T)$. It is possible to show that the social cost equals $2/(2+\mu T)\mu$.
\item {\bf Full control} The optimal social welfare for asymmetric strategies is obtained by one customer arriving at time~$0$ while the other arrives at time~$T$. The social cost is $\mathrm{e}^{-\mu T}/\mu$.
\end{itemize}
We can use the above to compute the PoA with respect to the optimal symmetric strategy: 
\[
PoA=\left\lbrace \begin{array}{ll}
1+\frac{\mu T}{2}, & T< \log 2/\mu,\\
\frac{2(2+\mu T)}{2+\mu T-\log 2}, & T> \log 2/\mu.

\end{array}\right.
\]
Similarly, the PoA bound corresponding to the full control solution can be derived. Interestingly, it is shown in~\cite{HO2018} that the symmetric equilibrium strategy under the processor-sharing regime coincides with the socially optimal symmetric strategy. This result does not extend to the case of more than two customers seeking service along the time interval~$[0,T]$. \hfill $\Diamond$
 
Going beyond the simple case of  two customers there is little hope for finding explicit solutions, as was the case for the equilibrium analysis. Therefore, the majority of research on social welfare optimization relies on algorithmic methods and approximations. For the ?/M/1 queue with only waiting costs and $\mathcal{T}=[0,T]$ \cite{HK2011} use a local search method on a discretized time grid to approximate the optimal symmetric arrival strategy. The approximated optimal symmetric strategy is quite intuitive: a mass of customers should arrive at the opening, a mass at closing and almost uniformly along $(0,T)$. It was further shown numerically that the gain in social welfare by not allowing early birds is very small for a system that is not very heavily loaded. The methodology of \cite{HK2011} relies on techniques used in outpatient scheduling, namely optimization of multimodular functions. See Altman et. al \cite{AGH2000} for theoretical background and Kaandorp and Koole \cite{KK2007} for the application to outpatient scheduling. 

In the observable model of \cite{BO2018} some interesting asymptotic properties of the PoA are established. For the full control benchmark it is shown that the PoA approaches 2 as the population size or waiting cost increases to infinity. When the PoA is defined with the optimal symmetric strategy then the same result holds for the population size only. The PoA approaches a much lower number ($\approx 1.06$) when the waiting cost increases to infinity.  

The task of optimal scheduling or sequencing of jobs has naturally received a lot of attention in the operations research literature (see Pinedo~\cite{book_P2005} for an extensive introduction to the topic) and some of these methods are also useful for the when-to-arrive model. Choosing the socially optimal  arrival times (full control) for a singe server with exponential service is shown to be a convex problem in Pegden and Rosenshine~\cite{PR1990}, when the objective is minimizing a weighted combination of expected waiting times and makespan. Hence, finding the optimal schedule can be achieved efficiently using standard methods such as gradient descent. However, for more elaborate systems the objective function is non-convex and sometimes it is hard to even evaluate the cost  for a given schedule. Thus, heuristic methods are often used. One such heuristic is the  equidistant schedule of arrivals spread out on $[0,T]$ (see, for example, Stein and C\^{o}t{\'e}~\cite{SC1994}). This heuristic has a solid intuition behind it that a social planner would like to spread arrivals as far as possible apart in order to avoid congestion. For the loss system of \cite{HR2015} it is shown that such a schedule is in fact optimal in the case of a single server and no queue buffer, but suboptimal for other systems. For more elaborate loss systems the equidistant schedule is used as a lower bound for the optimal social welfare and PoA. Together with  an upper bound obtained by an oracle upper bound (that knows all service time realizations) the heuristic is shown to perform well. Furthermore, this is used to show that the PoA is not monotone with the population size; it is low for small populations because customers spread out by themselves in equilibrium, low for large populations because the system is very congested for any schedule when the admission interval is relatively shot and so there is little room for improvement, but for moderately sized populations there is a big difference between the equilibrium and socially optimal outcome.

Generally speaking, the problem of social welfare optimization has less structure than that of deriving the equilibrium arrival process for which we saw there is a common framework in many of the papers. This means that typically adhoc analysis is required. For the model with tardiness and earliness penalties, but no waiting costs, \cite{GHR2017} show that  the socially optimal solution is an equilibrium (for most parameter sets), but it is not a unique equilibrium and so the price of anarchy is still greater than one.  Ravner and Nazarathy \cite{RN2017} consider the socially optimal schedule for the deterministic processor model of \cite{RHV2016}. The complex dynamics yield a piecewise convex objective that can be maximized using an exhaustive algorithm that is only feasible for a small number of customers. Heuristic methods are developed for the general problem. Numerical analysis in \cite{RHV2016} suggests that in equilibrium the deviation from the due dates is small in comparison with the socially optimal solution but congestion levels are much higher.

\section{Fluid approximations}\label{sec:fluid}

Fluid models with determinsitc dynamics are widely used to approximate queueing systems because they enable explicit derivation of performance measures and optimal policies. This is especially true for time-dependent systems that cannot rely on steady-state results (see, for example, Mandelbaum et. al \cite{MMR1998}). The fluid approach has also been heavily used in the transportation literature in the analysis of variations of the Vickrey bottleneck  model (see \cite{V1969,ADL1993}). Using our notation, the Vickrey model assumes the $(\alpha,\beta,\gamma)$ linear cost structure for waiting, tardiness and earliness costs, respectively. Furthermore, as the motivation typically comes from traffic networks composed of roads, it is assumed that the server is always available. In the stochastic queueing setting  similar assumptions to the ones made in~\cite{SK2017} (which was discussed in Section~\ref{sec:cost}) are made. The latter also compared the fluid approximation of the equilibrium profile to the stochastic counterpart without the restrictive assumptions often made in the transportation literature (e.g. $\beta>\alpha>\gamma$).

Fluid approximations in the context of when-to-arrive decision model were first introduced by Jain et. al~\cite{JJS2011}. They turn out to be much easier to analyze (as expected), are shown numerically to give good approximations and also lead to quick approximations to the price of anarchy (PoA). Juneja and Shimkin~\cite{JS2013} show that the equilibrium of the stochastic system converges to the fluid solution as $N\to\infty$. This result is unique in the sense that most papers first assume a fluid model, sometimes also proving that the discrete system actually converges in distribution to the fluid system (e.g., Honnappa and Jain~\cite{HJ2015} and Honnappa et. al~\cite{HJW2015}), and then derive the equilibrium of the game based on the fluid model.

Following a similar path to that of Section~\ref{sec:equilibrium}, we first provide a detailed description of the equilibrium solution for Example~3, which considers homogeneous customers with linear waiting and tardiness costs. Detailed analysis of this example appears in \cite{JJS2011}, \cite{H2013} and \cite{JS2013}. This will be followed by a review of additional literature that applied fluid models for approximating the equilibrium and socially optimal solutions.

\subsection{Fluid model of Example~3 (waiting and tardniess costs)}\label{sec:fluid_tardiness}

Fluid models make each customer tiny which can be considered as a drop in the ocean. Each drop is served in no time. The original volume of the drops is fixed to some non-random value $\Delta$, and the server can serve $\mu$ of them per unit time. In particular, working continuously the server can serve all demand in $\Delta/\mu$ units of time. Also, a drop that finds a volume of $\delta$ upon its arrival waits $\delta/\mu$ units of time prior to service commencement (which is also its time of departure). An intuitive explanation for the approximation of the stochastic system with a Poisson population and exponential service times is as follows. Suppose that $\lambda^{(m)}$ and $\mu^{(m)}$ are increasing and unbounded sequences of arrival and service rates such that $\lambda^{(m)}/\mu^{(m)}=\Delta/\mu$ for all $m\geq 1$, then as $m\to\infty$ we have an accelerated version of the stochastic system such that there is a huge population of infinitesimally small customers that can all be processed in $\Delta/\mu$ units of time. Of course, there is no need to assume an exponential service times to obtain the fluid limit.

The waiting and tardiness parameters are $\alpha$ and $\beta$, respectively, as before. Each drop needs to decide when to arrive along the time interval $\mathcal{T}=(\infty,T]$ (randomization allowed) and in this setting too a strategy defines a distribution over the possible arrival instants. Recall that the server starts serving customers only at $t=0$. Moreover, a symmetric Nash equilibrium is such that given that all drops follow it, this is also a best response for an individual drop. When it comes to social optimality, we assume that a central planner can dictate the arrival time distribution which, due to the features of the fluid approximation model, is equivalent to enforcing a specific (different) arrival
 instant on all. In other words, there is no distinction between the full  and the anonymous controls.
 
The first and biggest step in moving from the stochastic queueing model to the fluid model is in observing that the amount of work in the system in the latter case is a deterministic process. If the system is not empty during the arrival interval of customers $[0,T]$, which is obviously true for any equilibrium solution, then the queuing dynamics of \eqref{eq:E_Qt} become now
\[
Q(t)=\Delta F(t)-\mu t,\quad \forall t\in [0,T],
\]
for any symmetric arrival distribution. Furthermore, the waiting time for a drop arriving at $t$ is given by the deterministic process $Q(t)/\mu$. In particular, recall that the equilibrium density given in \eqref{eq:exmp3_f_equil} depends on the probability of an idle server $p_0(t)$. For the fluid system $p_0(t)=0$ for all $t\in[0,T]$ and so the equilibrium condition simplifies greatly as there is no need to solve elaborate dynamics such as the backward equations of the stochastic system. The equilibrium arrival distribution turns out to be uniform, possibly with multiple intervals with different densities.

Let us first consider the model where $T<\Delta/\mu$ and early birds are allowed. We provide a detailed outline of the analysis for this model and then summarize the results along with the results of the other three variations of the two above assumptions ($T\geq \Delta/\mu$ and no early birds, etc.). All cases can be handled in a similar manner and details can be found in~\cite{H2013}. For simplicity, we assume below that both cost parameters, $\alpha$ and $\beta$ are positive. 

Let $[-t_a, 0]$ be the early bird period in equilibrium for some to-be-determined $t_a>0$. For this time interval we have that the cost of arriving at $t$ equals
\[
-\alpha t + (\alpha + \beta)\Delta F(t)/\mu,
\]
which is constant in equilibrium. For $t\geq 0$ the cost is 
\[
\beta t + (\alpha + \beta)(\Delta F(t) - \mu t)/\mu,
\]
which equals the same constant. Thus, the equilibrium arrival distribution is uniform with density 
\[
f(t)=\frac{\mu}{\Delta}\frac{\alpha}{\alpha+\beta}, \ t\in[t_a,T] .
\]
The condition $\int_{t_a}^Tf(t)\diff t=1$ further implies $t_a =\frac{\Delta}{\mu}\frac{\alpha+\beta}{\alpha}-T$. 

The disutility of the one who arrives at time $-t_a$, which equals the disutility of all others, is therefore $\alpha \frac{\Delta}{\mu}\frac{\alpha+\beta}{\alpha}-T$. Hence, the social cost
in equilibrium is
\begin{equation}\label{SCEQEB}
\Delta(\Delta(\alpha+\beta)-\alpha\mu T)/\mu.
\end{equation}
As the server operates continuously from $t=0$ until the system is empty the social planner has no discretion over overall tardiness costs and therefore the socially optimal strategy minimizes the waiting times by equating the arrival rate and the service rate for as long as possible, i.e., on the interval $[0,T]$. Hence, the unique socially optimal arrival strategy is a uniform density of $\mu/\Delta$ along the interval $[0,T]$ and an atom of size $1-\mu T/\Delta$ at $T$. The corresponding social cost equals 
\begin{equation}\label{SCSOEB}
\frac{\mu\beta}{2}T^2+(\Delta-\mu T)(\beta T+ \frac{\Delta-\mu T}{2\mu}(\alpha+\beta)).	
\end{equation}
Those who arrive during $[0,T]$, a mass of $\m T$ , do not wait at all and, on average, each one of them incurs a tardiness cost of $\beta T /2$. This contributes a total of $\beta \mu T^2/2$ to the social cost. The rest, a mass of $\Delta - \mu T$, arrive at $T$. Each one of them already incurs a tardiness cost of $\beta T$. On average, they clear the system after an additional time of $(\Delta-\mu T )/2 \mu$ with a cost of $\alpha + \beta$ per unit time of delay.

Note that the PoA equals the ratio between~(\ref{SCEQEB}) and~(\ref{SCSOEB}). It is possible to see that when $\alpha = 0$ the PoA equals $2$. Moreover,
it increases with $\alpha$ and approaches the limit of $\Delta/(\Delta - \mu T )$ (which is not a function of $\beta$) when $\alpha\to\infty$. Furthermore, the PoA goes to $2$ when $T$ goes to zero or when it goes to $\Delta/\mu$. Also note that both the length of the arrival interval and the density along it are not functions of $T$: Shrinking the arrival period after opening leads to drops arriving earlier by the corresponding value. Finally and as expected, the length of actual arrival period depends on the values of $\alpha$ and $\beta$ only through their ratio.

We next summarize the results for all cases.  It is clear from the above analysis that if $T \geq \D/\m$ then in fact there is no limitation and in equilibrium as well as under social optimality customers anyway arrive  prior to the closing time $T$.

\begin{enumerate}
\item {\em The case with early birds and $T\leq \Delta/\mu$.} \\
If $T \leq \D/\m$ then the unique equilibrium arrival strategy implies a density of  $\frac{\m}{\D}\frac{\a}{\a+\b}$ along the time interval $[-\frac{\D}{\m}\frac{\a+\b}{\a}+T,T]$. The corresponding social cost equals $\D(\D(\a+\b)-\a\m T)/\m$. The unique socially optimal arrival strategy implies a uniform density of $\m/\D$ along the time interval $[0,T]$ and an atom of size $1-\m T/\D$ at $T$. The corresponding social cost equals $\frac{\m\b}{2}T^2+(\D-\m T)(\b T+ \frac{\D-\m T}{2\m}(\a+\b))$.
\item {\em The case with early birds and $T> \Delta/\mu$.} \\
The unique equilibrium arrival strategy implies a uniform density of $\frac{\mu}{\Delta}\frac{\alpha}{\alpha+\beta}$ along the time interval $[-\frac{\Delta\beta}{\mu\alpha},\frac{\Delta}{\mu}]$. The corresponding social cost equals $\Delta^2\beta/\mu$. The unique socially optimal strategy implies a uniform density of $\frac{\mu}{\Delta}$ along the time interval $[0,\Delta/\mu]$ with a social costs of $\Delta^2\beta/2\mu$. In particular, the PoA equals $2$. This result appears in~\cite{JJS2011}.
\item {\em The case without early birds.} \\
Here we deal with the same fluid approximation model with one key difference:
The seniority of those who arrive prior to the opening time is not kept. In particular, all those who arrive earlier enter service at random order. Thus, there is no point of arriving before $t=0$. We next give the details of the equilibrium strategy, distinguishing between $T \geq \D/\m$ and $T \leq \D/\m$. As in the discrete case, the corresponding distribution comes with an atom at zero and then a gap with no arrivals until some time $t'$. From then on until the last instant of possible arrivals, and as in the case with early birds, the density is uniform. Finally, as of course arriving prior to opening was not part of the socially optimal arrival strategy when it was allowed, the strategies detailed in the previous cases are also socially optimal in this model. Finding the PoA is a simple exercise. Details are given next for the two possible case.
\begin{enumerate}
\item {\em The case of $T> \Delta/\mu$.}\\
 If $\b \geq \a$, the equilibrium strategy is pure and it prescribes arriving at the opening. The corresponding social cost is
$\D^2(\a+\b)/(2\m)$ and the $PoA=(\a+\b)/\b<2$. If $\b \leq \alpha$ the equilibrium strategy is to arrive at zero with probability $2\b/(\a+\b)$, then there is a zero density (a gap) along the time interval $(0,\frac{\D\b}{\a\m})$
and a uniform density of $\frac{\m}{\D}\frac{\a}{\a+\b}$ in the time interval
$[\frac{\D\b}{\a\m},\frac{\D}{\m}]$. The corresponding social cost equals $\D^2\b/\m$ and the PoA equals $2$.
\item {\em The case of $T\leq  \Delta/\mu$.}\\ 
In the case where $\b \geq \a$ the behavior is as described in (a). Assume next that $\b < \a$. If $T \leq \frac{\a+\b}{2\a}\frac{\D}{\m}$, then the equilibrium strategy is pure: all arrive at time zero. The resulting social cost equals $\D^2\frac{\a+\b}{2\m}$. Dividing this value by
(\ref{SCSOEB}) yields the PoA.
If $\frac{\a+\b}{2\a}\frac{\D}{\m} < T < \frac{\D}{\m}$, then the equilibrium is with an atom of size $p=2(1-\frac{\a \m T}{\D(\a+\b)})$ at zero, a zero density (a gap) along the time interval $(0,t')$ where 
$t'=\frac{p\D(\a+\b)}{2\a\m}=\frac{\D(\a+\b)}{\a\m}-T$
and a uniform density of $(1-p)/(T-t')$ along the interval $[t',T]$.
As it turns out the corresponding social cost coincides with the one given
in~(\ref{SCEQEB}). Likewise, the resulting PoA is as in the case with early birds.
\end{enumerate}
\end{enumerate}

\subsection{Overview of fluid models}\label{sec:fluid_models}

As we have seen, the fluid dynamics enable explicit derivation of the equilibrium and socially optimal arrival distributions. This can be further used to study more elaborate systems for which even numerical methods for the computation of the equilibrium, such as Algorithm 1, are out of reach. The fluid solution was also used as a benchmark solution in many of the papers dealing with the discrete stochastic model that we have discussed so far (e.g., \cite{H2013,JS2013,HR2015,SK2017,RS2020}). Roughly speaking, the fluid equilibrium solution is indeed a good approximation for large systems with small customers, but for small population games the equilibrium densities can be far from uniform.

In many applications customers are not homogeneous in their utility functions or their service demand. The models we have considered so far have all assumed homogeneous customers, with the exception of \cite{TMP2019} that assumes that customers have a random waiting cost that is determined by a common distribution. Assuming that there are multiple classes of customer types using the same server makes the equilibrium analysis much harder. Consider Example~3 once more with $P$ customer types that have different linear waiting and tardiness costs $(\alpha_i,\beta_i)$ for $i=1,\ldots,P$. Every type of customers now have different equilibrium conditions of the form of \eqref{eq:exmp3_equil} and \eqref{eq:exmp3_equil2} that keep the cost constant throughout the arrival support. Hence, an equilibrium arrival profile is a solution of a system of functional differential equations that all have the same underlying backward equations of the common queueing process. Studying the properties of such a solution, e.g. existence and uniqueness, or even constructing a numerical method that always yields a solution is a challenging task and we are not aware of significance results in these directions. Some of the difficulties associated with multiclass when-to-arrive games are discussed in \cite{RS2020} for a model with service rate uncertainty. The fluid model enables analysis of the equilibrium in multiclass games, as well as other more complex models such as non-FCFS service regime and queueing networks.

The fluid version of the concert queueing game (Example 3) is studied in \cite{JJS2011}. The equilibrium arrival profile dictates that different types of customers arrive on disjoint time intervals with uniform densities that are determined by applying the same type of computations used in Section \ref{sec:fluid_tardiness}. This is further used to obtain lower and upper bounds on the PoA. An extension of the concert queueing game with multiple customer types, time-dependent service rate and non-linear tardiness costs is considered in Juneja and Shimkin~\cite{JS2018}. The form of the equilibrium is similar with disjoint arrival intervals for the different types of customers but the equilibrium dynamics are more elaborate, and thus the arrivals are not necessarily uniformly distributed. It is further shown in \cite{JS2018} that a unique equilibrium exists and an efficient method to compute it was presented. Juneja et. al~\cite{JRS2012} assume the fluid population volume is random (with a single type of customers and constant capacity). In this case explicit analysis is possible but the resulting equilibrium arrival density is uniform only on part of the support and decreases for values above some threshold.

So far we have dealt only with single-queue systems, i.e., even if there are multiple servers, there is a common waiting buffer for all of them. The when-to-arrive game for a network of fluid queues is considered by Honnappa and Jain~\cite{HJ2015}. Customers have linear waiting and tardiness costs as before, but in a network setting there is additional question of routing. It is assumed that routing is decentralized and so customers can choose both their arrival times and what queue to join. First, the equilibrium arrival profile for a network of $K$ parallel queues is derived. The equilibrium arrival profile dictates that the flow of arrivals to each queue $k\in\{1,\ldots,K\}$ is uniform on an interval $[t_k,T]$, where $t_k<T$ for all $k$ with $t_k<0$ for at least one $k$. The PoA is further derived, similarly to Example~3, and is shown to be bounded by 2. Moreover, the model is extended to multiple customer types that differ in the parameters of their linear cost functions. As before, for the multiclass model in equilibrium the different types arrive in specific queues on disjoint intervals. Finally, it is shown that certain more elaborate network topologies, such as queues in tandem with a bottleneck structure, can be studied by showing that they have an equivalent parallel network representation. Therefore, the results on parallel networks can be applied directly to these more elaborate networks. 

In Section \ref{sec:other_dynamics} we discussed the difficulty in analyzing stochastic discrete systems with non-FCFS service regime. However, these systems are common in many applications and can potentially improve social welfare (as we have already discussed for several examples). A comparisson of work conserving service regimes in a fluid setting appears in~\cite{PO2017}, with emphasis on FCFS and LCFS. Their model assumes a fixed (population) volume of customers that are restricted to arriving on $[0,T]$, linear waiting time costs, and general tardiness costs. It is shown that a unique equilibrium arrival distribution exists for both FCFS and LCFS. The LCFS equilibrium yields higher social welfare than the FCFS regime, and in fact FCFS yields the lowest social welfare among all work-conserving service regimes.

A when-to-arrive game for a fluid processor sharing system with capacity $\mu$ and a volume of $\Delta$ arrivals is studied in Juneja and Raheja~\cite{JR2014}. Note that processor sharing is equivalent in this setting to random order of service. Customers have linear waiting and tardiness costs as in Example~3. Clearly, in such a system customers  never choose to be early birds and therefore the arrivals start appearing at $t=0$, and with  an atom at zero. If there are no arrival time restrictions, it is shown there that the equilibrium distribution is given by an atom of size $\beta\Delta/\mu$ at time zero and a uniform density on $(0,\Delta/\mu]$. The case of a finite closing time is also detailed and follows similar arguments as in the corresponding case analyzed in Section~\ref{sec:fluid_tardiness}.

An implicit assumption of all the models in this review was that customers have the ability to arrive exactly at their chosen arrival time. This is of course not the case in many applications, for examples when going to the bank or a concert there may be a strictly positive variance in the travel time from home (or work, etc.). This phenomenon was modeled by Ghazanfari et. al~\cite{GLRN2021} for a single server fluid queue with customers that have linear waiting, tardiness and earliness costs. Specifically, customers choose their intended arrival time $t$ but their actual arrival time in the system is $t+X$ where $X$ is a random variable. It is shown that if $X$ has a uniform distribution, then there exists no equilibrium with a continuous mixed strategy. Furthermore, neither does a symmetric pure strategy equilibrium exist. Hence, a symmetric equilibrium has to take the form of a distribution with several atoms, which is very different from all of the models considered in this survey. The non-existence result for continuous distributions extends to other non-uniform distributions of the arrival time distortion $X$. An alternative dynamic agent-based logit decision model is suggested in \cite{GLRN2021} and is shown to converge to a stationary distribution for some parameters of the problem, where stationarity is defined in the regular sense: if customers arrive according to the stationary distribution $f$ on day $d\in\{1,2,\ldots\}$ then the logit induced distribution of arrival times on day $d+1$ is also $f$.

\section{Empirical results}\label{sec:empirical}
The motivation for the theoretical models surveyed here comes from decision making that is made in many daily activities that involve congested systems, therefore it is natural to ask how well does the theoretical Nash equilibrium predict the actual decision making and arrival patterns of customers. While the typical equilibrium solution is a mixed strategy, one does not expect that customers really flip coins before deciding when to join a queue, but rather that the aggregate distribution of the arrival times of all customers is well approximated by the equilibrium arrival distribution. In this section we discuss a stream of literature that set out to answer these questions by performing laboratory experiments that simulate the interaction of customers in a queueing system. The typical setup of these experiments is a sample of university students who interact anonymously in a game that mimics a queueing scenario. The participants are presented with a description of the system and the payoff structure and choose when to arrive to the queue. The realizations of waiting times and associated costs are translated to monetary gain (or loss) that participants receive at the end of the experiment.

The first experiment of this type was conducted by by Rapoport et. al~\cite{RSPS2004}. They introduced a discrete-time variant of the ?/D/1 model; $n$ customers choose when to arrive, from a discrete grid of time slots, to a single sever queue operating during an interval $[0,T]$ ( early birds were not allowed). The objective of the customers is to minimize their own waiting times. Multiple customers arriving in the time slot are randomly ordered. The symmetric mixed strategy Nash equilibrium is computed numerically and has the same form as the continuous-time solution of the ?/M/1 game as described in~\cite{GH1983}. Of course, as the discrete grid becomes finer, the games are essentially the same. Note, however, that in the discrete-time game non-symmetric equilibria may exist. In particular, for certain parameters it is possible that all customers arrive in different time slots and a queue is never formed (due to the deterministic service times). In practice such an outcome requires coordination between the customers which is not reasonable in most scenarios, especially when $n$ is large. The experiment was conducted for different parameter settings and four groups of 20 participants repeated the interaction multiple times. The main result of \cite{RSPS2004} is that although the individual decisions did not resemble the equilibrium strategies, the aggregate behaviour of participants in the experiments was remarkably similar to the theoretical equilibrium prediction. Specifically, many customers chose to arrive at the opening, the period immediately after the opening was rarely chosen, and then an almost uniform arrival rate throughout the rest of the opening period until closing. Similar results were obtained in Seale et. al~\cite{SPSR2005} for the game with early birds. The latter also conducted an experiment where the arrival time choices of all participants was made available to everyone after every repetition of the game, making it a repeated game, in which case the aggregate behaviour converges to equilibrium much faster. Similar experiments with similar conclusions were later conducted for a system with batch service in \cite{SRSZZ2007} and \cite{RSMZS2010}, with the latter considering the case of a random batch size. When there was uncertainty regarding the batch size, i.e., only the distribution for the batch is known, the aggregate behaviour did not display a similar pattern as the equilibrium prediction.

Breinbjerg et. al \cite{BSO2016} conducted an experiment that tested the arrival time decisions of customers to a single server system under three different service regimes: FCFS, LCFS and random order (RO). As before, for practical purposes the game was played on a discrete grid of time slots. The equilibrium analysis and experiments were confined to three-player games. The small-scale systems enabled computation of the theoretical equilibrium arrival distributions for LCFS and RO service regimes, which otherwise may be intractable as was discussed in previous sections. The experiments led to several interesting observations. First of all, the theoretical equilibrium did not display a good fit to the aggregate behaviour of the participants in any of the settings. The main difference is that in the experiments the arrivals were more spread out over the grid with less congestion in the opening. This was especially visible in the FCFS and RO regimes, where there is a strong incentive to arrive early. Interestingly, this disparity results in higher social welfare than predicted by equilibrium. Although there was no coordination between participants in the experiments, their arrival time choices can be figuratively described as somewhere in the middle between selfish individual optimization and coordination due to a centralized mechanism that minimizes expected overall congestion costs. 


\section{Discussion and open challenges}\label{sec:discussion}

We have surveyed the growing body of literature dealing with arrival processes to stochastic queueing systems that arise from strategic timing decisions of customers. The focus was on the theoretical foundations of such models. In particular, we have discussed the construction of appropriate mathematical models and the resulting properties of their solution concepts such as Nash equilibrium and social welfare optimization. The motivation for the study of such models comes from many daily activities that involve congested systems. However this research stream has so far devoted limited attention to the empirical study of real systems. Experimental laboratory studies have examined the decision making of uncoordinated customers in congested service systems and have found that the ?/M/1 model predicts the aggregate empirical distribution pretty well in certain cases, but there is still a disparity for some system settings. It would also be interesting to explore whether arrival data from actual service systems (i.e., not laboratory experiments) can be fitted with the ?/M/1 model. Suppose for example that data is collected daily on the arrival process to a system that has certain opening hours. The mean population size can be easily estimated by taking the average daily total number of arrivals. The next step is to estimate the cost parameters of customers, e.g., waiting cost $\alpha$ and tardiness cost $\beta$, as in  Example~3. This can be achieved by evaluating the empirical arrival rate at different times of the day and selecting as estimators the $\alpha$ and $\beta$ that minimize some loss function with respect to the distance between the equilibrium prediction of the arrival rate and the empirical observations. Of course, there is also a model selection problem here and for specific applications different models will have a better fit than others. Developing statistical methodology seems like a natural research direction that complements the economic analysis that has been reviewed here. 

There are certainly still many open questions in the theoretical analysis of when-to-arrive games. This is especially true for more complex systems with various service regimes, heterogeneous customers or networks of queues. As we have seen, the direct analysis of the stochastic system is very challenging and new methods are called for in order to make progress in this direction. Furthermore, much insight has been gained by using fluid approximations, however, in many cases the prediction of fluid models seems unsatisfactory. For example, in the multiclass concert queueing game of \cite{JJS2011} customers arrive at uniform rates on disjoint intervals but this type of behavior is not plausible for many systems. As was detailed in Section \ref{sec:fluid_models}, extending this fluid model to allow for non-linear cost functions, random population size and network structure yields equilibrium solutions with a more elaborate structure. Still a gap remains in the equilibrium analysis of small scale multiclass systems with atmoic customers. One approach that has potential to provide insight on the more elaborate stochastic systems is the dynamic learning model of \cite{SMF2019} discussed in Section \ref{sec:other_dynamics}. In terms of computation, one only needs to be able to simulate the dynamics of the system in order to compute the equilibrium approximation. However, there are no theoretical guarantees yet that the learning dynamics converge to an empirical distribution that is indeed a good approximation of the equilibrium solution. Developing a rigorous framework for the learning dynamics, including asymptotic analysis and error bounds on the approximation, would be very useful. This may be lead to a powerful tool for approximating the equilibria in complex systems that are otherwise intractable.


\section*{Acknowledgments}\label{sec:acknowledgments}
The authors wish to thank Refael Hassin, Yoni Nazarathy and an anonymous reviewer for their helpful comments. The first author acknowledges the financial support of the Israel Science Foundation, grant no. 1512/19.

\end{document}